# MICHAEL GURSTEIN

# WHAT IS COMMUNITY INFORMATICS
(and Why Does It Matter)?

Polimetrica

PUBLISHING STUDIES

directed by Giandomenico Sica

VOLUME 2

**MICHAEL GURSTEIN**

# WHAT IS COMMUNITY INFORMATICS
**(and Why Does It Matter)?**

Polimetrica

**Note for the Reader**

In our view, doing research means building new knowledge, setting new questions, trying to find new answers, assembling and dismantling frames of interpretation of reality.

**Do you want to participate actively in our research activities?**

**Submit new questions!**

Send an email to the address **questions@polimetrica.org** and include in the message your list of questions related to the subject of this book.

Your questions can be published in the next edition of the book, together with the author's answers.

**Please do it.**

**This operation only takes you a few minutes but it is very important for us, in order to develop the contents of this research.**

Thank you very much for your help and cooperation!

We're open to discuss further collaborations and proposals.
If you have any idea, please contact us at the following address:

*Editorial office*
*POLIMETRICA*
*Corso Milano 26*
*20052 Monza MI Italy*
*Phone: ++39.039.2301829*
*E-mail: info@polimetrica.org*

**We are looking forward to getting in touch with you.**

# LIST OF QUESTIONS

# INTRODUCTION

This brief volume is meant as an Introduction to my thinking about Community Informatics (CI) rather than specifically as an Introduction to CI. I make that distinction because at this stage in its early development CI represents a number of different things to different people.

To some, it represents a way of talking or thinking about a particular set of Information and Communications Technology (ICT) tools that are available for use in and by local communities. For others, CI is a form or methodology of Community Development that happens to use ICTs rather than blackboards as a primary means for facilitating community communications. For others, CI is a way of formulating and integrating the use of ICTs as an instrument for economic and social development into more mainstream Information Systems thinking and research. For still others, CI is the beginnings of a "movement" by means of which ICTs are appropriated by the marginalized to realize a new role for themselves in the Information Society. For myself, CI is a bit of all of these, but at its most basic CI is about a new but necessary way of approaching Information Systems and in fact represents an evolutionary advance on traditional systems by integrating them with the dynamism and adaptability of life as lived in organic communities.

The Question and Answer format suggested by the publisher is an interesting one in that it suggests a degree of informality and iterative thinking which is rather in keeping with CI at least in its current formulation. Also, a format such as this indicates that what is being presented is itself partial and subject to evolution and change along with the contents of what is being discussed.

A number of the specific areas presented are in fact adapted from other of the things that I have written in some case in collaboration with other people. I've tried to indicate where this has occurred but my apologies if any of these instances have been overlooked and my gratitude to my colleagues in each case where the development of the thinking was formally shared. Of course, all of the thinking concerning CI has been collaborative and iterative and thus all of this document should be seen as informally the result of the range of collaborations which I've been privileged to participate in over the years.

I would like to point particularly to my work with Richard Civille and our collaboration on our work for the Ford Foundation, Tom Horan for the paper we did together for the Davis Minneapolis workshop, Wal Taylor for continuing interactions in various parts of the world, and the multitude of colleagues who have contributed to the Community Informatics Research Network and the CRACIN e-lists.

The errors and omissions are of course, my own.

And of course, my thanks always to my wife Fernande Faulkner without continuing collaborations none of this would likely have ever seen the light of day.

Michael Gurstein (E-mail: gurstein@gmail.com)
Centre for Community Informatics Research,
Training and Development
Vancouver – Canada
http://www.communityinformatics.net

## *What is Community Informatics?*[1,2]

Community Informatics (CI) is the application of information and communications technology (ICT) to enable and empower community processes. The objective of CI is to use ICT to enable the achievement of community objectives including overcoming "digital divides" both within and between communities. But CI also goes beyond discussions of the "Digital Divide" to examine how and under what conditions ICT access can be made usable and useful to the range of excluded populations and communities and particularly to support local economic development, social justice, and political empowerment using the Internet.

CI is emerging as the framework for systematically approaching Information Systems from a "community" perspective and parallels Management Information Systems (MIS) in the development of strategies and techniques for managing their community use and application. As well, it is closely linked with the variety of Community Networking[3] research and applications.

CI is based on the assumption that geographic communities (also known as "physical" or "geo-local" communities) have characteristics, requirements, and opportunities that require different strategies for ICT intervention and development from the widely accepted, implied models of individual or in-home computer/Internet access and use. Also, CI addresses the concern for ICT use in Developing Countries as well as among the poor, the marginalized, the elderly, or those living in remote locations in Developed Countries.

CI represents an area of interest both to ICT practitioners and academic researchers and to all those with an interest in community-based information technologies. CI addresses the connections between the academic theory and research, and the policy and pragmatic issues arising from the tens of thousands of Community Networks, Community Technology Centres, Telecentres, Community Communications Centres, and Telecottages currently in place globally.

What characterizes a CI approach to public computing is a commitment to universality of technology-enabled opportunity including to the disadvantaged; a recognition that the "lived physical community" is at the very center of individual and family well-being – economic, political, and cultural; a belief that this can be enhanced through the judicious use of ICT; a sophisticated user-focused understanding of Information technology; and applied social leadership, entrepreneurship and creativity.

CI presents a challenge to technology and technologists to respond to the needs of those who wish to use technology to better their daily lives. Equally CI presents a challenge to academics and academic researchers to move beyond narrow and self-reflexive research disciplines towards an engagement with the ways in which technology is and can be made useful in transforming living conditions and life chances in both the developed and developing world. Finally, CI presents to policy makers the need to recognize the ways in which ICT is now making possible new modes of governance, self-organization and self-management.

A theory and a practice of Community Informatics is gradually developing. Partly this is arising out of experiences with community access and community networks and partly out of a need to develop systematic approaches to some of the challenges which ICT is surfacing with astonishing speed, including the recognition that access in itself is

insufficient – rather it is what is and can be done with the access that makes ICT meaningful. CI is also developing to ensure a local, civic and "public" presence in an increasingly commercialized Internet environment and as a basis for enabling and supporting local innovation particularly as this is requiring an adjustment to technology change and the globalization of production and competition.

CI from my perspective has multiple and intersecting antecedents. These include:

- Organizational studies of the application and implementation of information systems;
- Management Information Systems as the study of the design and development of systems in support of organizational goals;
- Social activism as the undertaking to accomplish certain normative objectives in the world (in this case through the use of ICT);
- Community development as the process of animating and bringing communities to a state where they are able to self-manage and self-organize;
- Policy studies and public administration as the understanding of the means by which the public interest may be realized through the implementation of public programs;
- ICT for Development (and development studies) as the use of ICT as a tool to promote social and economic development; and
- Service design as the approach to determining how services may be delivered most efficiently and effec-

tively (in this case in relation to community goals and through the use now, of ICT).

All of these come from different perspectives and with different histories and modes of implementation. In fact, it is through the process of implementation and practice that all of these have become fused as CI activities; and along with the social processes of which they are a part, they have come to interact with and respond to the range of social environments for ICT implementation and have elicited a variety of local responses and adaptations which are the heart of Community Informatics.[4]

*What role does CI play in an Information Society/ Networked Society?*

**KEYWORDS:**
INFORMATION SOCIETY, MANAGEMENT INFORMATION SYSTEMS, NETWORKED SOCIETIES, MANUELL CASTELLS, NETWORK, COMMUNITY INFORMATICS, ICT

The development of the Information Society (or probably more appropriately individual national "Information Societies" is also a massive restructuring of various societal processes, structures and institutions as these are driven to adapt to the opportunities (and risks) presented by digital technologies. However, this restructuring does not deploy in a purely random or disinterested fashion. In fact, the development of these new societal structures conventionally operates so as to enrich and empower those with the resources (and the capabilities) to manage and direct those processes in their own interests both on corporate and individual levels.

As a typical example, Management Information Systems (MIS) are conventionally developed precisely as a way of supporting (enabling/empowering) management and corpo-rate processes (and not as would otherwise be possible, processes in support of labour, consumer, or other "stakeholder" interests). These systems provide management with the capability of controlling and directing resources including of course information resources through the use of information technology. The significance of this is often lost since the usual way of approaching information systems is as though these were neutral or disinterested rather than as the matrix through which power is deployed within institutional structures. The significance of Community Informatics in this context is that CI is a basis for empowering those in society who are normally without power in these systems.

It might thus be more useful and revealing to be discussing the role of ICT in "networked societies" rather than "information societies", as the former would focus attention more on the manner in which ICT is precipitating this restructuring of economic and social (and political) systems (and including the power relations within these) rather than on the content of such systems i.e. "information" where such restructuring (and its impact on power distribution) may be less immediately visible. In this we are of course following Castells in his highlighting of "networks" as the dominant and pervasive structural feature of the digitized society and economy although other of our observations and conclusions in this regard diverge rather directly from Castells' positions.[5]

What we mean here by the "networked society" is a society where the emerging if not (yet) the dominant structure of productive relations is one where production (particularly information based production) is undertaken primarily by means of and transacted through a technology (ICT) mediated infrastructure. In this context then the role of the network, networked relationships and the instance that such relationships are necessarily ICT-mediated, has a pervasive transfor-mative effect on all aspects of these network-based transactions and interactions.

The "Networked Society" is a society which is re-forming itself based on open-ended and widely dispersed network structures and connections rather than, for example, on those structures which reflect the requirement for face to face connections or physically mediated relationships such as families. In this sense, the notion of the *network* is to some degree in opposition to the notion of *community* in that networks are conventionally structured around the relationships of autonomous and self-directed individual actors (or nodes) where the basic structuring is of individuals (nodes) interacting with other individuals (nodes) with linkages between nodes being based on individual choice. The contrast here with communities is that *communities* assume collectivity or communality within a shared framework which may include common values, norms, rules of behaviour, goals and so on.

The role of communities in the Networked Society is in this sense, to provide creative and dialectical tension with and a counter-post to the individual-based "network" from the perspective of (and to achieve the benefits of) broader sharing and collectivity. An analysis or an "informatics" based on "communities" thus is in contrast to the dominant mode of analysis which sees technology mediated networks as consisting of inter-linked individuals aggregated only through their positioning as quasi-independent nodes.

In the current and conventional approach to the Networked Society, the contrast is sometimes presented as "networks" being **technologically** enabled with "communities" being based on **social and human** connections. In fact though, in some if not in many cases information technology can act equally as the enabler of communities as with the enablers of networked individuals/nodes. This process of enabling communities (through Community Informatics) thus becomes a process of community informaticization with the information technology acting as the carrier and the facilitator of the connections either or both between communities or within communities. These communities in turn may be formed with equal efficiency and effectiveness on an IT platform, with ICT (as with non-ICT-mediated connections) in this instance, acting as a community enabler alongside its more readily recognized role as the enabler of individual-based networks.

## *Is Community Informatics a Critique of or Parallel to Networked Individualism and the Facebook Society?*[6]

**KEYWORDS:**
BARRY WELLMAN, NETWORKED INDIVIDUALISM, SOCIAL NETWORKING SOFTWARE, NETWORK, IDENTITY, SELF-ORGANIZED NETWORKS, ONTOLOGY, COMMUNITIES

Barry Wellman and his colleagues[7] have presciently characterized the nature of the status and relationships between individuals within electronic networks as being that of "networked individualism".[8] An example of this is of course the type of "networking" which takes place within the rapidly emerging social networking software category (viz. Facebook, Myspace, Flickr etc.). In addition, although to my knowledge they haven't made this connection, networked individualism can be seen as the most appropriate means for conceptually framing the emergent form of employer-employee relationship most clearly characterized by Wal-Mart's "associates". Here rather than managers and management organizations providing the basis of work coordination and organization it is the electronic infrastructure, the "network" which provides this coordination and structuring of work activities including the integration of individual work activities into the broader corporate and administrative structure.[9] This employment structure is based on a form of individualism quite unknown in earlier management employee-relationships. The reason that this is possible, and the basis for its occurrence and the strength of these structures in relation to individual employees is, of course, the pervasive availability and application of ICT.

Wellman's notion of "networked individualism" as the way in which identity manifests itself in the "networked" society is a useful one in that it highlights both the manner in which the network links into the individual in a socially unmediated fashion (the network is the social mediation rather than for example family, ethnic or religious group or community) and the manner in which the individual both experiences and interacts with the dispersed and (from his/her) perspective centreless network directly, rather than through the mediation of any of these social groupings. It also gives a sense that the individual, in the context of an environment where she is engaged in multiple electronically-enabled connections ("networks"), is in turn a construct linking together fragmented identities/individualisms structured or created in response to the variety of only partially, if at all, overlapping, networks. Thus the creation of the self in this context may be (and is generally) understood as an act of individual will which may take different forms for different individuals or even different forms for the same individual in different contexts or occasions.[10]

Following from this, it should be understood that individual action within this context takes place not in a social framework but rather within and through the individual networks where the self is able (or available) to act (or interact) with others, but only within the very limited areas of linkage or interconnection that are presented by the specific network. As an example, an individual game player, part of a Massive Multi-Player Online Role-Playing Game (MMORPG) is able to interact with her fellows and act within the game but only in ways and with actions that are prescribed and circumscribed within the specific parameters of the network and thus the rules (and defined identities) of the game. Similarly the individual buying or selling on E-Bay, performs their respective actions within the parameters defined by the interaction between the individuals as per their "profile" within E-Bay and within E-Bay's prescribed and technologically enforced rules of interaction or "policies".[11]

The notion of identity and through this individual action as a "networked individual" is thus a peculiar one in that while the individual may define their specific "identity" within the context of the specific network (the definition of the individual's "profile" within that network), the manner in which that identity may in turn execute or perform actions within that network are directly a function of the centrally determined and prescribed standards or regulations, i.e. the "code" of that network.[12] The individual may control their profile (identity) but they can do so only within rules over which they have no direct influence and which they can resist or ignore only at the risk of being de-networked and thus de-linked or *erased* from participating in the network. In network terms, of course, being erased is tantamount to being obliterated – i.e. not simply killed or destroyed, in which case traces may be allowed to remain within the network – but obliterated i.e. removed including all historical traces or fragments.[13]

Within Wellman's model of "networked individualism", the only ontological mover (independent agent or source of independent action/agency) is the *network* itself. The *individual* in Wellman's formulation is simply the sum of the fragments of her participation in the various externally driven networks (of production, consumption and even socialization) of which she is a member or with which she has contractual relations. In this world, the *network* is all and everything.

However, in the real world, the externally driven network is only one element of reality. In addition to this, there are the self-initiated (self-organized) and participatory networks which inter-link individuals not on the basis of fragments of identity but on the basis of self-initiated and self-realized identities. These networks function as "communities" (whether based on physical or virtual connections) through which action may be undertaken, projects realized, reality confronted and modified.

These "communities" both physical and electronically-enabled, represent an additional (and structurally oppositional) ontology to the "network" ontology as described by Wellman. These communities provide a basis or a foundation element for the construction of an alternative reality. This alternative "reality" is in practice a set of organizational, economic and social structures which operate independently of the centrally controlled networks and are capable of opposing and creating different processes, structures and "realities" to those being produced (and forcefully reproduced and extended) through the centralized/individualized networks as discussed by Wellman and as realized by such corporate agents as Wal-Mart and (as will be discussed below) Microsoft.

This observation, that "communities", whether electronically- or physically-enabled have an independent and self-standing ontological status is a significant one since within much of the social science and technical literature, "communities" are regarded as phenomena secondary to either individuals, groups or corporations. The assertion here is that "communities" can and should be seen as free-standing and primordial and as the platform or conceptual agent on the basis of which for example, one could (and should) undertake technical i.e. hardware and software, design.[14] In this way, one can (and this is the conceptual foundation for a "Community" Informatics) specifically develop information, communications and networking systems which would provide the means for communities to be enabled and empowered and to effect action in the world. This approach would be directly parallel to the use of information, communication and networking technology for enabling and empowering "corporations" (or as in the social networking context, "individuals").[15]

*What is the relationship between networked (virtual) Community Informatics and (place based) Community Informatics?*

#### KEYWORDS:
NETWORKED COMMUNITIES, BOTTOM-UP, USERS, PARTICIPATION, COMMUNITIES, COLLABORATIVE DECISION MAKING, EMERGENT COMMUNITIES, SUSTAINABILITY, VIRTUAL OR ELECTRONIC COMMUNITIES, PEER TO PEER, CONVERGENCE

"Networked communities" (equally with community networks) have a variety of "essential" characteristics which differentiate them from other and centrally determined networks and networked individuals.

Networked communities are "bottom-up", that is they are developed and driven by the users or participants themselves rather than being centrally initiated or externally driven. What this means is that users or participants are actors in the networks and these networks in turn are community-based, developing through pre-existing or self presented individualisms rather than (as discussed above) with the inter-individual connections being externally defined and elicited. In this way participation in a community is rather more rounded and integrated from the participant's perspective than the fragmented and largely "contractual" or rule based relationships of the individualism based networks as defined by Wellman. This in turn gives the nature of the community participation a stronger and fuller grounding in the lived context.

Participation in networked communities is voluntary and self-initiated. Individuals participate in networked communities on a voluntary basis, that is they choose to participate (in the case of physical communities it is rather in the form of choosing not to "not participate") and this participation is based on individual decision and volition rather than through entering into contractual relationships. There may be an exchange of "value" (in the form of cash, goods or services) through a networked community, in fact there is very often quite a considerable such exchange, but that is not the basis of the participation in the community and the relationships are with the community as a whole rather than on a bi-lateral basis where there is an enforced or enforceable structured value exchange relationship.

In communities, the goals and the methods for achieving these goals, are the result of collaborative decision making processes. These processes of course, may differ significantly from context to context but in each case there is an element of participation by those involved and responsiveness to the decisions made. In practice these processes reflect some form of "consensus" position on the part of the participants although the achievement of formal consensus may or may not occur and in many communities there are many, more formalized, structures for decision making.

Networked community structures are autonomous and capable of the independent initiation of action. In this context, networked communities function as the "edge" of the various larger networks in which they are participants. As in the Internet itself, the notion is that the intelligence (and relatedly the capacity for autonomous action and independent i.e. non-coerced participation in the network) is found at the edges of the network.

Networked communities are "emergent" in that they come into existence (rather than having a formal substantive reality over time), often in response to some external condition or circumstance. That they are "emergent" doesn't mean that they don't or haven't persisted for a lengthy period but rather that they may have lain nascent until called forward into formalized existence by an external stimulus or by internal processes of (for example) "social entrepreneurship" or self-initiated problem-solving. In the same manner "networked communities" may and do evolve over time and move into and out of "existence" in formalized terms. That no formal structures of a "networked community" can be externally identified does not mean that "the networked community" does not exist, but rather it is that the networked community in its structured form is still nascent waiting to be called forth.

This approach provides a means to understand the "sustainability" paradox which is that while the formal structures of communities may or may not be "sustainable"[16] over time, nevertheless the "community" itself is sustaining and may "spring to life" i.e. re-emerge in the form of formalized structures at a future but as yet unpredictable occasion. This suggests the obvious but frequently overlooked conclusion that "communities" are not defined simply by their structures, but rather are the connections which persist over time as between members of the community, with structures being simply formalizations of these connections.

Communities in a networked environment (including networked communities) may take either of two forms: they may be physical communities which are enabled both internally and in their relationship with the outside world through the use of ICT or they may be communities which only exist in and through the electronic networks which enable them.

In the latter instance, these communities may also be known as "virtual" or "electronic" communities indicating their origins in the act of networking and of inter-individual peer-to-peer communication. In many cases these communities reflect a repurposing of top-down centrally driven e-networks where individuals as end users/participants in these networks begin to by-pass the central network authority and enter into direct peer to peer communication. Centrally driven networks are almost always structured so as to preclude the possibility of peer-to-peer connections (by-passing the centre) recognizing that this type of "organizing" would be of little advantage to those at the centre (and in control) and could potentially present threats.[17,18]

Further, as the use of ICT to support electronically enabled communities becomes common, and as experience in enabling physical communities with ICT is acquired there is emerging a convergence or an over-lap between these two forms of community. Thus for example, electronically enabled communities are beginning to seek out ways of becoming linked more directly into physical interactions and physical processes and ICT enabled physical communities are beginning to enhance and extend their activities and reach by incorporating elements of virtual relationships as aspects of their on-going physical and face-to-face relationships.

This on-going trend towards convergence is built on a recognition of the value-added both by the virtual connections (information at a distance and access anytime, anywhere) and of face to face connections (the opportunity to rapidly build trust and to overcome the ambiguities and distortions of computer mediated communication). With both of these presumably sharing the longer term goal of an optimal seamless inter-connection and inter-weaving of the physical with the virtual and the enormous power for the achievement of collaboratively identified outcomes.

*What is the relationship between Community Informatics and Social Activism?*

**KEYWORDS:**
SOCIAL POWER, INFORMATION SOCIETY, GLOBALIZATION, WAL-MART, RESISTANCE, PLACE-BASED COMMUNITIES, CODE, PEER TO PEER, INDIVIDUALISM, PROFILE, COMMUNITY, RESISTANCE, ACTIVISM

There is an assumption in CI that ICT are the new source of social power and that access to and use of ICT is the basis on which power is being redistributed within society. Community Informatics at its core is thus about power and how power is distributed in society.

As we drive forward into the "Information Society" the overwhelming force of globalization not simply as a metaphor but as a defining condition of the dominant structures of the emerging economy becomes increasingly evident. Globalization in this context means the creation of centrally coordinated networks of producers and consumers, of supply chains and distribution networks. And necessarily and crucially these processes in their current late $20^{th}$ and early $21^{st}$ century manifestations are enabled and empowered by ICT.[19] The very rapid rise to national and increasingly global dominance of a select number of massively electronically enabled corporations of which the most visible and successful is Wal-Mart (in the retail sector) is the defining example of these processes.

What is characteristic of Wal-Mart and all of the companies linked into the Wal-Mart web of alliances, suppliers and sub-suppliers as well as the companies whose own drive for globalization emulates or parallels that of Wal-Mart is the very high degree of centralization and centralized control which they exert even through their highly dispersed operations.[20,21]

A notable characteristic of Wal-Mart as well, is not only is it a "globalized" enterprise but it is also a "globalizing" enterprise and not only in the reach of its business activities but also and even more tellingly in the drive of its technology systems to create ever more efficient structures for information flow and information management along its supply and sales chains; its extended supplier eco-system; and of course towards an ever-smoother integration of the two.[22]

And yet, we are seeing manifestations of resistance and even coordinated resistance to these networks and their impacts and particularly in the context of physical communities throughout the US and elsewhere. Notably, the only effective resistance to the Wal-Mart juggernaut and including competitive resistance in the market place has come initially from place-based communities and in general, relatively integrated small communities which have mounted active resistance to the location of a Wal-Mart store within their immediate environment even where for example, by contrast a variety of unions have spent very large amounts of energy and money attempting to organize Wal-Mart employees with notably little success.[23]

It should be noted that in this context, the capacity to resist or to "organize" within the network is in itself a feature determined and circumscribed by the network's code. This code in turn determines in what manner "individuals" are allowed to interact and coordinate their behavior outside of the centrally prescribed coordination as for example, determined by the rules and standards of the individual networks.

Only by stepping outside of these networks and drawing upon or creating an "individualism" which is either outside of these networks or based on a non-network derived identity is collaborative action or "organizing" possible. Otherwise, this type of interaction (i.e. *peer to peer*) is precluded through the subordination of inter-individual connections to those which are allowed within the structure of coordinated networked relationships. In environments or for individuals whose identities are largely structured in relation to these externally driven networks – those for whom employment, gaming, purchasing networks – are the sum of their "individualism", little may be left as a residual base of identity on which to form such non-externally coordinated inter-relationships (i.e. non-externally dependent networks) as a foundation for *resistance*.

The electronically enabled network that is Wal-Mart and the parallel centrally-controlled electronically-enabled networks that underpin the contemporary advanced economies and cultures are, at their very core, "totalizing" systems in the sense in which thinkers such as Hegel and Marx understood "totalizing systems". That is, they are systems whose inner life is one of extreme and even cancerous and explosive growth and through this to the absorption or transformation of ever wider circles of production (and consumption) into extensions of these ever-expanding network chains.

And remembering that the characteristic mode of human participation in these networks is through a necessarily fragmented participation as an individualistic networked electronic "profile", it is not surprising that the resistance to this totalization comes from opportunities and frameworks which enable the individual to overcome this fragmentation and to integrate their identity and more importantly to find the means for entering into collaborative relationships with others. This process of re-integration or overcoming contractually structured and fragmented "networked" relations in favor of organic and wholistic coordinated relationships is in fact what takes place in "communities" and is in some senses the defining characteristic of communities – where *Communities* are places "where others know your 'name' and not just your 'sig' (electronic signature) and where others interact with you as an integrated person not simply as an electronically mediated 'profile'".

Thus the structure of "resistance" to the totalizing forces of technology and network-enabled market accumulation as per Wal-Mart is necessarily and theoretically (as well as in practice) the discovery or rediscovery of "community" and of organic and integrated inter-individual relationships rather than purely contractual and electronically fragmented inter-networked connections.[24]

And these relationships of resistance can be seen not simply as resistances in themselves, but also as dialectically produced and structured resistance (in struggle) with the networked structures being invasively engendered by the Wal-Mart and similarly structured technology platforms. In fact, the relationship here is a "dialectical" relationship between "community" and centrally-controlled, distributed electronic networks which are the characteristic form that capitalist production is currently taking.

Thus CI as a vehicle for enabling and empowering is at its very core an "activism" or at least an enabler of activism in that it is giving the tools of empowerment to those whose need for such instruments it greatest.

*What role does Research Play in CI Development?*

**KEYWORDS:**
RESEARCH AND DEVELOPMENT, MIS, NETWORKED COMMUNITIES, COMMUNITY-BASED SYSTEMS

Research has played a very significant role in the advance of centrally facilitated networking. The development of the technical systems including the basic network infrastructure, the hardware on which it resides, and the software which enables it are all the products of research and development ac-tivities and to a considerable degree of university supported engineering and computer science research.

The basic paradigm of this research has been that of the design and development of MIS, that is, information systems designed so as to enable concentrated and centralized decision-making by (primarily) corporate management. In addition these systems are used as a means for achieving efficiencies in business applications and increasingly for managing and controlling the various inter-relationships which corporations have with those outside the corporation with whom they have on-going relationships whether as customers, suppliers, lenders, regulators and others.

The research driving much of this advance has of course, been technical in nature. The basic managerial assumptions (centralization of decision making, structuring around a "command and control" model, exclusionary access to information and so on) on which this technical research is founded are deeply embedded and taken as necessary and commonplace. Any suggestion that these assumptions are anything but necessary is viewed with suspicion and incomprehension. Even the research which is concerned with the application and use of Information Systems in not-for-profit enterprises makes the same basic assumptions concerning the nature and structuring of the organization and the necessity of empowering centralized management.

However, the research issues and strategies for networked communities are somewhat more complex than for managed networks in that networked communities require that the assumptions and characteristics of themselves as communities must be included as elements and assumptions within the context of the research. Thus, the research component for CI requires an understanding and sensitivity towards the *social* elements of computing and communication in addition to the *technical* elements. As well, there is a need to break away from received research designs and assumptions given that networked communities must be seen at some levels as being in a *dialectical relationship* with the dominant, intrusive and engulfing forces of the centrally-coordinated networks and networking.

In this context then, *research* is not simply about the nature of the activity or how it might be enabled using ICT. It must also be informed as to the nature of the on-going and structured relationships between the research and how it supports the use of ICT to enable dominant social and economic forces (or alternatively the relationships of resistance or *struggle* which are necessary responses to the exertion of this ICT enabled power). It is through the development of the latter forms of community-based systems where CI makes its contribution, i.e. through ensuring the provision of the ICT functionality and of the enabling and empowering ICT systems and associated services.

*What is the difference between Management (Information) Systems research (MIS) and Community Information (Informatics) Systems research?*[25]

**KEYWORDS:**
COMMUNITY INFORMATICS RESEARCH, MIS RESEARCH, APPLICATION, COMMUNITIES, PRACTICE, OUTCOMES, INFORMATION SYSTEM, SYSTEMS DESIGN, COLLABORATIVE DECISION MAKING, ACTION RESEARCH, "LOCAL" VS. "GLOBAL", COLLABORATIVE SYSTEMS, INTELLIGENCE AT THE EDGES, EFFECTIVE USE

Community Informatics research is necessarily wholistic including paying specific attention to and being explicit concerning the particularities of the social context in which CI systems are to be implemented. This is in contrast to MIS research where such assumptions are buried and implicit and it is assumed that findings or recommendations are universal and context independent.

Also, for CI the goals of the application are at least in part linked directly into the requirements of the networked community through the complexity of responding to the emergent needs of communities in their diversity and the specificity of their functioning as communities. This is in contrast to MIS applications where goals are relatively straightforward in the sense of being concerned with ensuring efficiency of operation and thus profit maximization. CI goals are necessarily more diffuse and of course, include the various specific elements required to reflect and ensure the continuity of the "community" for which the application is being developed.

This "wholistic" approach requires therefore, a much greater understanding of and insight into the broader social and even physical (as well as technological) environment within which the application is being introduced. It follows from this that the nature of CI is primarily "practice" i.e. outcome driven, rather than methods or theory driven.

For MIS the link to outcomes is in relation to business (or organizational) applications or business practice and the ultimate measure of success is the degree to which the outcomes are usable and provide useful results. For CI, the practice is parallel to this but here of course, the measure of success is the degree of usefulness or usability by end user communities. As above, given the emergent and impermanent nature of networked communities, there is a clear need for continuous and structured linkages and feedback (and feed forward) mechanisms between researchers, practitioners and community users.

A central organizing notion in this discussion is the question of who is to be identified as the "owner" or operative "agent" for an Information System (IS). For classical IS, the agent is clearly the "organization" or more particularly the "manager" as in *Management* Information Systems. But in the context of community-based applications of IS whether in the form of community technology centres or electronically enabled communities of practice, the operative agent is clearly not a "manager" or "managers" or even an "organization" as such. Rather, the "owner" of the system and ultimately the target user from a design perspective would be a somewhat more dispersed or "democratic" (consensus driven) operative agent such as a "community" (understood here as a collectivity of individuals linked around a common goal).

In this latter context, systems design rather than being premised on providing support for what are in effect "management" decisions by organizational agents, would come to be framed around what are (to the community) internal and dispersed and/or collaborative (even consensus driven) decision-making processes. Further, an equally common operative or guiding design principle (even a tacit assumption) for IS i.e. that of promoting or enhancing the "efficiency" of an operation, may be of less significance (or in practice subject to contestation) in a community context where principles of design stressing inclusiveness, representativeness or equal access may be of equal or greater importance. Thus for example, an IS approach to the design of an e-Government system might look to optimize speed of transaction processing and efficiency of process management while a community-driven (CI) e-Governance system might look to maximize inclusiveness, transparency and dispersed and democratic participation in service governance and delivery.

From a research perspective, the organizing notion for CI is that of an inter-disciplinary approach to communities and technology, combining computing and engineering with social science; community, social and economic development; and social practice. In addition, CI incorporates as a necessary element of its system design strategy, approaches that are consis-tent with and contributory to the methodology of Action Research as a way of knowing for Information Systems.

The contention here is that CI provides a fundamentally different set of organizing principles and processes that is best considered a cousin to, rather than a derivative from traditional MIS, namely concerning:

- *Inputs*. The MIS aim is to process and provide information that can improve the efficiency and profitability of an enterprise. In a CI system a primary input for system design is the social or community context of the system – who the users are, what they need, what skills they bring (or don't bring) to the use of the system. The CI system then could be said to be designed around and for the users and secondarily for the processes that are being supported.

- *System goals*. Local is everything in CI. CI systems act so as to enable the "local" rather than the "global", the "dispersed" rather than the "concentrated", the "decentralized" rather than the "centralized" or the "disaggregated" rather than the "aggregated". Thus the basic model of information technology is not one of "hub and spoke" where the technology core (application, information source, service provider, and transaction hub) is at a "center". Rather it is one that favors "peer-to-peer" or face-to-face communication where the self-selected participants are randomly dispersed but in loose electronic connection with each other through the core/center or "peer-to-peer" relationships.

- *Process*. In classic MIS, the focus is on using such systems to deliver competitive advantage to firms through actionable information on business activities. CI can be seen as favoring the "communal" that is the "collaborative' rather than the "competitive" and thus it supports open systems and the free sharing of knowledge; collaborative or consensus rather than hierarchical decision-making; and inclusive rather than exclusionary or selective decisions or information sharing processes.

- *Outputs*. For CI, the information or service outputs are strongly determined by the enacted views of those stakeholders that use, want, or value social and community related information. Thus a CI system is one where output, intelligence about the output, and the means to influence output is distributed to the edges rather than concentrated in a center or hub. This intelligence and the information that is the resource that enables its use, is thus available to communities and users for the achievement of their goals and not, as traditionally, concentrated in a standardized model, such as a "scorecard" of outputs available only for use within a highly constrained and competitively focused context.

- *Outcomes*. CI favors "effective use" rather than "passive access"; "production" rather than "consumption"; and information or knowledge creation and use, as the basic model for end-user computing by those at the system periphery. Thus, rather than a concern with responding to, for example, the "Digital Divide" by extending opportunities for technology or Internet access, the issue is redefined as finding ways of using ICT's to respond to issues and requirements that are meaningful and significant in the daily lives of individual users within their communities. There is also a concern for empowering and enabling end-users (within community or collaborative contexts) to make use of the systems to support their applications and as a response to their needs.

- ***Contextual Factors***. As noted above, the system context in many ways drives the differences in inputs, process, and outputs. In classic MIS, there is a business to run and there is a need for information as an assist in improving competitive performance. In CI, the community is more fluid in terms of its direction and fundamentally responsive to the desire for and willingness to contribute information by users, community members and other stakeholders. Societal gains in terms of community growth, equity, and quality of life are much less susceptible to a priori codification (and thus translatable into highly structured systems). Thus the supporting systems are required to be much more responsive to emerging interests and needs and open to assisting communities-as-users in identifying and clarifying system goals appropriate for them and as well as the most effective systems applications for achieving these.

## *Are there specifically CI areas and principles of Research?*

**KEYWORDS:**

DIGITAL DIVIDE, ECONOMIC AND SOCIAL DEVELOPMENT, RESEARCH, SUSTAINABILITY, RESEARCH DESIGN, KNOWLEDGE SHARING, RESEARCH NETWORK, RESEARCH "PEERS", RESEARCH AS "PROCESS", TECHNOLOGY AS POWER, PARTNERSHIPS

Very considerable public (and private philanthropic as well as commercial) resources are being directed toward responding to the perceived Digital Divide as well as using ICT as a platform for enabling economic and social development among marginalized populations and in Less Developed

Countries. Determining how these resources might most effectively be deployed, as, for example, through "community" institutions, and the most appropriate strategies and models, requires both formalized research and the systematization of the range of practical experience.

The research issues of interest to CI include how communities can become the "subject" of technology applications and how technology in turn can enable communities to become more active, effective and secure as "subjects"; the differing strategies required for urban and rural, low and moderate income, digitally literate and non-literate communities to become technologically enabled; strategies for "re-engineering community processes" of environmental and land management, cultural production, and democratic participation/empowerment; appropriate and sustainable business models for community based e-commerce initiatives; and the most effective methods of scaling/linking these processes laterally between networks of similarly enabled communities. A further issue of considerable practical importance is the on-going economic/institutional "sustainability" of local access--how it will survive once initial funding sources and volunteer participation are exhausted. The issue of "sustainability", of course, raises issues of the on-going benefits that ICT provide to community members.

It is possible to identify some working principles or general guides to CI research practice:

1. **The use of the research is to be built into the research design itself**: CI research is not generally done simply for the research but rather in relation to a specific outcome or action in the world of practice.

2. **Building-in strategies for knowledge sharing and collaborative knowledge building**: A key element of CI research is the contribution that it makes to the larg-er CI research and practitioner community, thus an element of the research design must be the identification of a strategy for contributing to and participating in knowledge sharing and collaborative knowledge-building right from the beginning.

3. **Research knowledge as an output of the "network" rather than a solitary "hero"**: The basic model of CI research is that of the research network (including academic researchers, practitioners and those involved in policy – all of whom contribute to the research in their own way and all of whom derive specific benefits from the outcome of the research). The research model is thus not that of the solitary "hero" researcher gathering knowledge and bringing it forth in authoritative pronouncements to an expectant universe.

4. **Non-researchers as research "peers"**: Similarly, there is in CI research a recognition and an acceptance of non-researchers as research "peers" i.e. as equal partners in the design, conduct and analysis of research.

5. **Research as "process" rather than "product"**: CI research is an on-going and iterative engagement between the researcher and the "subject" of the research and thus moves back and forth in an iterative fashion between problem definition, information collection, analysis, engagement at the level of practice, assessment and feedback and then back again to problem definition and so on.

6. **The technology is an instrument of power**: CI-oriented technology is technology which enables communities to achieve a degree of persistence and a degree of autonomy in the midst of attempts to eliminate these zones of independence. Thus while individual items of

> technology may in themselves be seen as neutral for use by either side of these struggles, the broad force of the technology and thus the manner in which it is specifically instantiated is understood within CI as either supporting or undermining the autonomous and self-organizing role of networked communities.

The notion of "partnerships" between community users and researchers is a powerful one as well, but also one with some associated difficulties. Differences in short-term objectives (and criteria of achievement – users looking for applications, while researchers are requiring certain formalized institutional acknowledgement as for example that which results from peer review); differences in language and even in cultural norms and practices; incommensurable schedules and timelines (users looking for immediate results with researchers being to a considerable degree, governed by institutional commitments and calendars); and so on make the relationship (or working partnerships) between CI researchers and communities or practitioners somewhat difficult, but not impossible.

*What is the relationship between Community Informatics research and Community Informatics theory and practice?*

### Keywords:

THEORY, APPLIED RESEARCH, PRACTICE, COMMUNITY ONTOLOGY, EMERGENT PHENOMENA, MODELS, NETWORKED COMMUNITIES, ITERATIVE APPROACH, COMMUNITY PROCESSES, ACADEMIC DISCIPLINE

Theory in CI as in other areas of applied research has the role of informing and guiding practice, and of giving guidance to research in relation to practice. Specifically in CI research, theory is needed to provide insight into the particular areas where the community ontology presents design or application challenges which diverge from those which underlie other areas of applied technology. Thus for example, how can one understand, conceptualize and model dispersed and consensus based decision-making as a basis for collaborative action and as a design criterion for technology systems to enable such processes.

As well, the challenge of CI research to enable or empower "communities" as persistent formalized structures even though "communities" in this context may be neither permanent nor fully realized. Rather they can be seen as emergent phenomena in the context of responding to the presentation of opportunities (or threats) from the larger environment; where, notably, the centrally driven networks operate with a continuous drive to encroach on and engulf all areas of activity including whatever small areas of autonomous action communities might be able maintain within these pressures and encroachments.

The role and objective of CI research is to document (within the context of ICT) these areas of conflict and resistance; identify those areas of small victory (where autonomous community-enabling activities and objectives are realized); determine those strategies which have achieved success; and suggest means for replicating, reproducing and extending these. Additionally, the opportunity for appropriating, integrating and repurposing existing technology as community supports while equally facilitating the development of technologies which in their very design reflect the specific ontology of communities presents very significant challenges and opportunities for CI researchers.

In this context the deeper and more formalized understanding of collective and non-hierarchical decision-making and consensus-building and the effects of electronically me-diated communication on these, will all inform the CI research in its relation to the outcome as practice. In this as well, since there are deep interplays between the specific nature (affordances) of individual technologies and the related interactive processes, these understandings (formulations, "theories") are works in progress rather than once for all universal insights. These understandings as well, may as readily take the form of inductive constructs (models) as of deductive propositions leading to formalized conclusions.

And, as the ultimate test of theory in this context is its usefulness and appropriateness of fit in relation to on-going and evolving practice, theory itself has to be seen as an ongoing and evolving set of formulations. These formulations finally, can be seen as responding both to changes in the technology environment and to the specific situations/contexts of its application.

What CI can do is to provide a larger framework and context into which each of multiple individual initiatives might be placed locally, nationally, globally and in terms of the broad development of a Community Informatics theory and practice overall. Thus, rather than seeing the success or outcome of individual initiatives as being slight particularly when only looked at in their specific local environments; when seen in a different light, as outcomes of struggle with the truly massive forces of the centralized and encroaching technologies of enforced dependency (including of course, the technologies of e-Government and e-Service delivery) they can be reinterpreted as being real successes.

The networked community is by its nature iterative in that its nature changes – grows, evolves, shrinks, disappears – in a recurrent and responsive fashion. As well, the various instances of its substantiation (formalization) may, and very often do, grow from and build on one another and thus evolve along with the technologies on which they are based.

Thus networked communities (and community networks) change over time, and design and analytical processes applied to these equally must recognize and make provision for such iterations and evolutions.

What this means is that CI theory and practice is always partial, and temporal and even context-specific, providing insight and direction for future developments and activities but necessarily reflecting and representing reality as based on a recognition and (interpretive) understanding of the nature of the local social and technology context and how this mediates any development or application. Thus, as findings and insights are realized, they may have a value in guiding and informing future action, design and implementation. At no stage can it be taken as given that there is a CI theory or a practice research that is once for all or has identified approaches which are universally "necessary" rather than locally "contingent".

In addition, this type of iterative approach implies a specific relationship between research and practice (one of partnership and knowledge sharing) and a certain humility in the manner in which research is presented and reported. Finally, there is a necessary recognition that results will always be partial and in evolution and that ultimately their value will come from the insight they provide as a basis for future action rather than as a once-for-all development of universally applicable models or theories.

Community Informatics at its base is necessarily a practice. That is the basic motivation for working within a CI framework is to act in and on the world through integrating the use of ICT with community processes and within the context of community objectives. As well however, CI functions within an academic environment as a research discipline and as such provides the basis for professional training for those working, implementing, or managing community-based tech-nology initiatives. In both areas it is still emerging, with a number of recent initiatives in universities and colleges and with attempts in various parts of the world to provide formalization and certification for community based technology practitioners.

As an academic discipline CI draws resources and participants from a wide range of backgrounds including Computer Science, Management, Information and Library Science, Planning, Sociology, Education, Social Policy and Rural, Regional, and Development Studies. As a practice, CI is of interest to those concerned with Community and Local Economic Development both in Developing and Developed Countries and has close connections with those working in such areas as Community Development, Community Economic Development, Community Based Health Informatics, Adult and Continuing Education, and Agricultural Extension.

## *What is the difference between Community Informatics and Social Informatics?*

**KEYWORDS:**
COMMUNITY INFORMATICS, SOCIAL INFORMATICS, ROB KLING, SOCIETY, DESIGN, ACTIVISM, INFORMATICS

There has been some continuing discussion about the relationship between Community Informatics and Social Informatics. Social Informatics (SI) is generally linked with the work of Rob Kling and his colleagues at the Centre for Social Informatics at the University of Indiana. Kling defined "social informatics" as "the body of research and study that examines social aspects of computerization, including the roles of information technology in social and organizational change, the uses of information technologies in social contexts, and the ways that the social organization of information technologies is influenced by social forces and social practices."[26]

Based on this definition then CI differs from SI in that:

a. SI is specifically concerned with "research and study" while CI is concerned with the "practice" (as well as the research) of the use of ICT in a "social" context;

b. SI is concerned with the very general and indeed abstract category of "society" and "social" or "societal" "aspects of computerization" while CI is concerned about how ICT are used in specific concrete identifiable communities;

c. CI among other things is concerned with specific applications of ICT in social/community contexts (health, economic development, education) while SI is concerned at the more general social or organizational system level;

d. CI is of direct interest (and is directly interested in) the design and development of ICT hardware and software (as well as applications) while SI seems to have little direct interest in the design or development aspects of ICT; and

e. CI lends itself to an "activist" involvement (not simply studying but getting involved in changing the role and significance of ICT in the world) while SI is content with simply attempting to describe and understand.

The statement is sometimes made that CI is a "subset" of SI but I don't believe this to be true. The practice component of CI strongly differentiates the "problematique" being addressed by CI from that being addressed by SI even though there is to some degree an overlap in the "subject matter" or "research" terrain. There is of course, the possibility for con-fusion since an "informatics" approach to social structures or social processes would to some degree be parallel to the "informatics" approach to community structures and community processes and equally, there is a clear linkage and even nesting of community structures and community processes within these social structures and social processes. However, I think that there may in fact be a misuse of the term "informatics" as used in the phraseology of "social informatics".

In that sense, I have difficulty in understanding what a specifically "social informatics" application might be. I can understand a "social networking" application, or a social servicing application but a specifically "social informatics" application doesn't at least for me have a clear referent or identifiable example.

## *Is Community Informatics about anything more than the "Digital Divide"?*[27]

**KEYWORDS:**
DIGITAL DIVIDE, INTERNET, ACCESS, INFRASTRUCTURE, DEVELOPMENT, EFFECTIVE USE

A search on Google finds almost 1,000,000 entries for the Digital Divide (DD). Of these some 700,000 refer specifically to the U.S., and an almost equal number to Canada and 300,000 to the DD in Less Developed Countries. There are a variety of definitions of the DD of which that at the "Whatis" Web site is perhaps representative: "The term 'digital divide' describes the fact that the world can be divided into people who do and people who don't have access to – and the capability to use – modern information technology, such as the telephone, television, or the Internet. The digital divide exists between those in cities and those in rural areas. For example, a 1999 study showed that 86 percent of Internet delivery was to the 20 largest cities. The digital divide also exists between the educated and the uneducated, between economic classes, and, globally, between the more and less industrially developed nations".[28]

A further definition goes on: "One third of the world population has never made a telephone call. Seventy percent of the world's poor live in rural and remote areas, where access to information and communications technologies, even to a telephone, is often scarce. Most of the information exchanged over global networks such as the Internet is in English, the language of less than ten percent of the world's population. This 'digital divide' is, in effect, a reflection of existing broader socio-economic inequalities and can be characterized by insufficient infrastructure, high cost of access, inappropriate or weak policy regimes, inefficiencies in the provision of telecommunication networks and services, lack of locally created content, and uneven ability to derive economic and social benefits from information-intensive activities".[29]

What is generally not discussed in the many studies and commentaries on the DD is how proposed solutions to the DD "problem" or condition, i.e. "improved access", will, in fact, provide any sort of useful response particularly to the impact that the DD is having. The underlying reasons for the impacts of the DD such as on-going trends towards increasing social and economic polarization – with the well-off getting better off and those behind falling even further behind as they find themselves unable to take advantage of ICT opportunities – are largely ignored. What, for example, is the link between "access" and wealth creation/economic development and does simply providing "access" do anything to provide that "missing link"? Is it reasonable (or useful) to indicate the need for "access" without suggesting a parallel need for training for use; structured links between "access" and production or distribution systems; or targets of "access" which correspond to individual or community needs in usable formats; and so on?

Of course, "access" (to the Network, to I/O devices, to content) is fundamental and basic to all other developments and uses of ICT technology. Without "access" little else is possible. However, the nature of that "access" is not without ambiguity, whether for example, the concern is for simple "access" as, through multiple user environments such as telecentres or whether there is a concern to provide in-home "personal" access; and, what about the quantity, quality and format of that access – Broadband, WiFi, dial-up. Which "access" is sufficient to "bridge the DD" and how or when do we know this?

The tendency moreover is to understand "access" as "technical" or "infrastructure" – related particularly by those directly involved with the issue as for example, those working in "development" or more broadly with policy or regulation in Less Developed Countries (e.g. through telecommunications regulatory agencies or development funding or policies). The result is a greater awareness and capacity to respond to perceived failures in "infrastructure" than there is in other possible issues concerning "access".

However, the use and application of ICT as the basic instrumentalities of the "Information Society" go much beyond discussions of the DD. They include examining how and under what conditions ICT access can be made usable and useful i.e. how "effective use" can be achieved by, among others, marginal or excluded populations and communities. Developing strategies and applications for using ICT to support local economic development, social justice and political empowerment; ensuring local access to education and health services; enabling local control of information production and distribution; and, ensuring the survival and continuing vitality of indigenous cultures are among the most significant possible applications and goals.

While considerable development resources have been spent on creating ICT infrastructure and access points (e.g., local telecentres), few of these initiatives, have been directed towards expanding local capacity for developing, managing and maintaining ICT capabilities. Additionally, the kind of ICT developments which would enable an effective participation by local communities in regional, national and even global decision making processes (e-Governance) are largely ignored in favour of the design and implementation of efficient and ever more centrally controlled if electronically enhanced systems for (e-Government and e-Commerce) service delivery. Again the early promise of the Internet as a means for enabling widespread distribution of the means for effective active citizenship has not been carried forward.

*What is "Effective Use" and what is its role in Community Informatics?*

**KEYWORDS:**
ACCESS, INFRASTRUCTURE, EFFECTIVE USE, ICT, CONTENT, I/O DEVICES, SERVICE, DEVELOPMENT PROCESS

"Access" in the conventional context is about being able to consume and receive rather than produce and distribute. Participation in the "Information Society" as presented from an "access" perspective is concerned with the capacity to purchase, to download and to interact passively with one or another externally created Web site. Bridging this DD clearly has as its goal to ensure that everyone is accessible to consumer goods and electronically mediated marketplaces.

The difficulty with "access" as the primary concern for those looking to ensure socially equitable use of ICT are the questions identified by Clement and Shade – "access for what", "access for what purposes", "access for whom" and "access to what". Without attention being paid to these issues, "access" as most commonly presented within the context of the DD discussion is simply a matter of ensuring opportunities to passively "consume" Internet enabled services and Internet supplied goods or information.

The notions of the Internet as a productive tool (or more broadly as an instrument for transformative change), in fact as the central productive tool of the Information age and of economies whose basic platform are ICT, is lost. Being a "producer" in this context is reserved only for the few. In practice, this is understood as being only for those working for corporations or governments, technologically advanced nations and those communities with specific training and skills as might be required to produce (and not simply consume) in a technology environment. In these contexts the opportunity (and the benefits which follow) of being a producer as well as a consumer are thus available not for those who have simple "access" but for those in the privileged position of designing and developing the uses and applications to which this access is being put.

The key element in all of this is not "access" either to infrastructure or end user terminals (bridging the hardware "divide"), rather it is having the knowledge, skills, and supportive organizational and social structures to make effective use of that access and that e-technology to enable social and community objectives. And even within the context of the consumption of services and goods by means of ICT, without attention being paid to the manner in which access is provided, many if not most will not be able to take advantage of the benefits available because of design or other flaws.

ICT is different in that once available it can readily become not simply a means to deliver content but also a means for the production, distribution, and sale of "content" locally or globally; and a basic infrastructure for production, distribution, sales and service in any area with a significant information, knowledge or learning component. ICT, it should be clearly understood, are the "satanic mills" of the Age of Information, but contrary to those "satanic mills", ICT presents the opportunity for very widely dispersed application and use.

"Effective Use" of ICT might thus be defined as: *The capacity and opportunity to successfully integrate ICT into the accomplishment of self or collaboratively identified goals* and includes:

1. *Carriage facilities* – What telecommunications service infrastructure is needed to support the application being undertaken? What are the appropriate and required volumes and capacities of bandwidth to be provided by broadband, dial-up, WiFi, satellite or other networked telecommunications services? What will it take to ensure that a supportive technology infrastructure is available in the form and quality (bandwidth, error rates, etc.) necessary to accomplish the purpose to which it will be put?

2. *Input/output devices* – What are the devices which users need in order to undertake a particular activity or use? Are they for example, computers for information processing, Personal Digital Assistants (PDA) for mobile information access, printers for text production?

3. *Tools and supports* – What software, physical supports, protocols, service supports are required? For example, databases for keeping track of large volumes of environmental data will be needed by environmental management groups, while physical textbooks may be a requirement for effective use by teachers of the support facilities of Internet enabled educational systems.

4. *Content services* – What specifically designed content is needed for particular application areas? What are the usability and locally contextual requirements for this content – language, design, literacy level, localization of references, links and so on. Effective use implies content which is designed to be specifically "effective" – usable, trustworthy, and designed for particular types of end users in the appropriate language formats.

5. *Service access/provision* – What type of social and organizational infrastructure, links to local social networks, para-professionals, training facilities are necessary for the particular use being developed? Effective use for many application areas requires an enabling social, as well as technological, infrastructure. Thus for example, effective use of e-health services in remote areas will require not just the technical access to the physical infrastructure, the information, the I/O devices and the service design but it will also require health application infrastructures including health care providers, para-professionals, community support systems i.e. a social organizational structure for the service to link the information or service being provided into local organizational structures and related delivery and support systems.

6. *Social facilitation* – What local or regional authorities/resources, community and environmental infrastructure, training, animation are required to locally enable the desired application or use? The effective use of an ICT enabled service will frequently require supporting facilitation since the service likely will not be effectively implemented spontaneously. There will be the need for coordinated planning and design, for training at all levels and for animation of the supporting structures to make the service usable. Overall of course, there will be the need for local leadership.

7. *Governance* – What is the required financing, regulatory or policy regime, either for governance of the application or to enable the implementation of the application within the broader national legal or regulatory systems? In many cases effective use will require an enabling financial structure, a supportive (or at least not inhibiting) legal or regulatory system as well as political support. Thus, for example, a major restriction on the effective use of e-health services has been the failure of many pre-paid health systems, both private and governmental, to develop financing systems which allow for reimbursement of the cost of electronic health support services provided remotely to end users.

When we are referring to the notion of "effective use" we are significantly extending the focus beyond possible ICT "tools" for development to highlight the entire "development process" including the infrastructure, hardware, software, and social organizational elements that all must be combined for development to occur. Clearly "access" is a pre-condition of "effective use". However, "effective use" as a design and development parameter for ICT is not necessarily included in conventional approaches or understandings of responses to the DD.

Issues concerning "effective use" are moreover, significantly contextualized, that is, what is an effective use in one context will not necessarily be so in another context. The focus on "effective use" is on individual use, or the user, or user community. The opportunity for defining and developing strategies for "effective use" should become a dialogue between those responding to the perceived inequalities of the DD and end users who understand most clearly what applications or uses would be most beneficial in particular local contexts.

All of the above it should be noted is presented solely in a passive and analytic mode. In the real world, there is the need for active participation on the part of the local community to "animate" or "appropriate" the process of technology acquisition and implementation. There needs to be a community "pull" as well as or in advance of the "top down" or external "push". Even before this there is also the need to create the local "pull" since in many cases communities or local users will be unaware of what types of opportunities are available through ICT. Thus, there may be the need for local animation or community development at the very beginning of an effective use approach to ICT implementation.

*What are some typical examples of Community Informatics applications?*

**KEYWORDS:**
APPLICATIONS, SUSTAINABILITY, LOCAL ECONOMIC DEVELOPMENT, K‑NET, FIRST NATIONS, ICT INFRASTRUCTURE, SOCIAL NETWORKING, SERVICE DELIVERY, COMMUNITY BASED TECHNOLOGY, SELF‑MANAGEMENT

Community Informatics is more an approach to applications than a specific application itself. CI also understands that the process of introducing an application may ultimately be as important for long term value and sustainability as the specific content of the application. Thus for example, a community based (or CI) application for local economic development is concerned with developing at the local level the capacity to manage and deploy the range of resources needed to support the local economic development activity in addition to a supporting a specific, say e-commerce, initiative or investment.

Probably the best example of a fully formed CI application is that of K-Net in Northern Ontario, Canada.[30] K-Net is the technology (ICT) and telecommunications arm of the Federation of Northern Chiefs and is one of the largest and most significant First Nations (aboriginal) technology users and implementers in Canada and very likely the world. Beginning from a base as a telecommunications integrator and service provider K-Net has moved forward to develop and manage its own highly sophisticated and extended ICT infrastructure including broadband, satellite, dial-up and other communications facilities, all to support an increasing range of information/knowledge intensive services for the 27 communities (some 27,000 people) in Northern Ontario many of which are only accessible by air. The services include a K-Net sponsored and managed high school, training programs, distributed e-health and tele-health services, and a fascinating social networking service (providing social connections among mostly young people but also families and others throughout the region resulting in there being some 30,000 email addresses in the region, substantially more than the total numbers of people).[31]

What is particularly notable about K-Net is the degree to which service design, development, and implementation has been undertaken at the direction and under the control of the local community service provider (as represented by K-net) and the local communities themselves. Thus K-net has operated so as to adapt and develop the range of services using an ICT base but specifically concerned with making those services as useful and usable to the community as possible and as well, providing a range of supports and development benefits both direct (service delivery) and indirect (local employment creation, local expenditures, literacy upgrade, training). It should be noted that, not surprisingly, this position and the development and success of these services has not been without conflict as K-net has had to struggle with both the traditional communications service providers (to gain control over the communications infrastructure) and the content service providers (to shift the management and control of these services from centralized service providers to locally based services and service managers).

K-net is at one end of a lengthy spectrum on which many if not most community based technology initiatives may be found. An examination of how K-net has come into being and how it operates provides a useful perspective on the various dimensions through which other such initiatives may develop. The primary dimension and difference between K-net and other community based initiatives is the degree of self-management which they have achieved. As they often note in their discussions, a significant difference between they and other such groups and activities is that they have had the capacity to acquire the means to control their own information management and communications infrastructure. This gives them the opportunity to negotiate around the nature and distribution of the range of services coming into their communities, something that other such communities/projects have not yet achieved.

## *Is there a specifically Community Informatics model of e-service delivery?*

**KEYWORDS:**
E-SERVICE DELIVERY, LOCAL KNOWLEDGE, INDIGENOUS POPULATIONS, INFORMATION SOCIETY, ICT, EDUCATION, REMOTE MONITORING

In fact, there is a specifically CI model of e-service delivery and one could argue that this model is at the very core of the overall CI approach. This model is one that focuses on using ICT as a support to communities as they manage the delivery of services to themselves. The broad trend within an increasingly professionalized and service intensive economy is to transfer responsibility to specialized institutions whose loyalty and broad responsibility is to agencies, influences and interests much beyond those of local communities. Thus for example, education which at one time was a largely community responsibility concerned with the transfer of local knowledge and the integration of young people into on-going community processes and structures is in most environments now under the management and control of professional bodies – governments and professional educators – with little knowledge or interest in community processes or priorities. This transfer largely took place because of increasing complexity in the nature of the knowledge being transmitted and the increasing need for specialized knowledge and skills to transfer this knowledge to young people.

The transfer of responsibility and control over education to distant authorities without local knowledge has proven to be a problem for many and particularly for rural and remote and indigenous populations. The requirement for modern skills has meant in a number of cases that for young people to gain an education it was necessary for them to leave their homes and families and travel to and reside in distant locations where urban temptations outside of family structures has often meant personal difficulties or an overall rejection of education. The alternative of not leaving home in these instances has meant that whole areas of education and training have been unavailable to certain young people and hence they have had no access to the employment and self-development opportunities which are dependent on these. ICT however, has the capacity to provide equal access to knowledge and to whomever (or wherever) those looking either to acquire or to inculcate this may be located. Thus many of the information access reasons for the professionalization and de-localization of education no longer hold in a technologically enabled information society.

There remains of course, the difficulty that the local community may no longer or currently possess certain of the specialized content skills required for education in certain contemporary areas of learning. However, a CI approach would involve looking at the service, in this case education, as something that the community should be enabled to provide and where ICT can be the overall means by which the service could be delivered. The approach would involve re-thinking the service and shifting from a perspective where there was an overall trend towards increasing professionalization and specialization to one where the service was conceptualized as something to be community delivered but within a context where ICT were used as a means to deliver information and skill support to community based service providers.

The service would need in this instance, to be re-thought and redesigned so as to be accessible to being delivered by relatively less professionally skilled community service providers; and where the information communication, delivery and management capability of ICT was used as a means to provide support for this alternative approach. In the case of primary and even secondary education for example, there would be the opportunity to integrate local service delivery (through trained local para-professionals) with additional machine based diagnostic and analytic capability, the use of remote monitoring and mentoring, and the design of teaching modules so as to make maximum use of the affective qualities of local delivery systems combined with the maximum use of the analytic capabilities of appropriately designed information support systems. The use of such an approach at least for identified components and stages in the education process would overcome many of the difficulties experienced by those living in remote and rural communities while still ensuring an appropriate level of education for young people.

A similar approach could be conceived of for the delivery of a wide range of primary health care services where, rather than transferring responsibility for primary health care to professionals, much of the initial health care would be designed so as to be delivered by community based para-professionals appropriately trained and supported with applica-tion designed information systems and information content. Such an approach would both allow for an increase in the access at the community level to primary health care (the most significant component of the overall health care system by volume of activity) while drastically reducing costs by shifting the cost structure of this service area away from highly trained and costly professionals and the hospital based services on which they rely, to community based (and much cheaper para-professionals) supported by a range of monitoring, diagnostic, information and assessment tools provided through specially designed ICT.

## *What is the role of Community Informatics in ICT policy?*

**KEYWORDS:**
GOVERNMENT POLICY, TOP-DOWN DECISIONS, BOTTOM UP PROCESSES, CANADIAN COMMUNITY ACCESS PROGRAM, LOCAL COMMUNITIES, SUSTAINABILITY, ICT

Government policy and including ICT policy is generally concerned with the making and implementation of top-down decisions. In the case of ICT what this means is for example, the development of projects, programs, policies, standards, regulations and so on that are determined at a bureaucratic or governmental/political level based on criteria and advice that are directed towards general requirements and then applied (imposed) on to the local circumstances in a standardized fashion.

The notion of bottom up processes and particularly of bottom up processes enabled by ICT is one that is almost completely alien to government and government policy even in the context where it would seem most obviously applicable (as for example where programs are being implemented in support of local social or economic development). Thus, where there are such programs – funding decisions, infrastructure installations, and application requirements are determined on a "one size fits all" basis and specified by public servants who may, but mostly don't, have local knowledge or experience.

The Canadian Community Access Program which was designed specifically to tap into local communities and local community resources, for the longest time resisted attempts to create inter-community networks as a basis for mutual support and resource aggregation. As well, the attempts to more closely tie project resources to local requirements and local resource availability foundered on the shoals of administrative practice and regulatory requirements. The result was, as might be expected, very limited success for the program and very great difficulty in achieving any degree of local sustainability.

Parallel processes can be found wherever such programs and policy structures have been addressed to the ICT sphere and including, most destructively, in the provision of support for ICT in the context of development processes in Less Developed Countries. In those instances, often funded either by multilateral agencies or bilateral donors, the insistence on top down processes or rather the refusal to allow for bottom up participation in planning and decision making with respect to program policies and program design has resulted in widespread program failure and financial waste and, to a degree, the creation, among some, of an overall negative perception of the value that ICT can contribute to economic and social development.

In those few instances where communities have been allowed to integrate ICT at their own pace and within their own community cultural and decision making structures, the utilization has been significantly more successful including in the development of innovative applications and strategies for deployment and use. K-Net in Northern Ontario is one of the best examples of this but others can be found in the work of the MS Swaminathan Foundation in India and the Savodaya Foundation in Sri Lanka.

Thus the role of CI in relation to government policy is to provide a continuing pressure towards the development of policies and programs supportive of bottom up ICT development, implementation and use and to undertake research and evaluations, including developing models and strategies which counter the prevailing top down norms and directions.

*Is Community Informatics anything more than Community Networking or Telecentres and what of the role of Universities?*

#### Keywords:

COMMUNITY, NETWORK, COMMUNITY NETWORK, COMMUNITY NETWORKING, ASSOCIATION FOR COMMUNITY NETWORKING, COMMUNITY BASED ICT APPLICATIONS, COMMUNITY TECHNOLOGY CENTRES, TELECENTRES, CYBERCAFE, EFFECTIVE USE, ACCESS, E-GOVERNMENT SERVICES, SELF-MANAGEMENT, UNIVERSITIES AND COLLEGES

Merriam Webster dictionary defines community as a "unified body of individuals" or "people with common interests living in a particular area". A Merriam Webster definition for "network" is "a system of computers, terminals, and databases connected by communications lines." The combined definition could be: "A unified body of people with common interests using a system of computers, terminals, and databases connected by communications lines." A somewhat broader definition that includes the technical wording while incorporating social values derived from the above might be: *A community network is a locally based, locally driven communi-* *cation and information system designed to enhance community and enrich lives.*

The term "*community network*" or as process, "*community networking*" has been in common use by thousands of community-based ICT projects in many countries for many years, and combines the sense of both the geo-local and online contexts depending upon its usage. The Association for Community Networking, in its inaugural organizational publication defined "community networking" as occurring: "when people and organizations collaborate locally to solve problems and create opportunities, supported by appropriate information and communication systems. A Community Network is a locally-based, locally-driven communication and information system".[32]

However, as the cost of Internet access has declined, the primarily middle class (and university based) activists who created and maintained the community networks in the developed countries along with their users have tended to lose interest and most of what had been several hundred community networks globally have now disappeared. These have been replaced as the focal points for community based ICT applications (in marginalized communities in Developed Countries and more generally distributed in Less Developed Countries) by what are variously called community technology centres, community access points, telecentres and so on.

In the context of CI, community telecentres are currently the focal point through which ICT involvement in many if not most communities takes place. Telecentres often (particularly in less developed countries) are the site where ICTs and community organizations and community processes come together (through those organizations). Of course, in many communities, cybercafes, that is businesses offering Internet access to all comers for a fee may be in competition with telecentres or may in fact offer, the only Internet access in the community. The difference of course between a telecentre and a cybercafe is not simply that one is "free" and the other is fee-paying (some telecentres charge access or other fees) but rather that the telecentre is a site where activities, services, or targeted programs may be undertaken in support of community activities; while cybercafes are simply sites for individuals to interact with and through the Internet – the purposes of which (at least in theory) are known only to themselves.

This difference means that for example, telecentres are likely to have staff, funds, possibly software and other resource materials all available for supporting various kinds of ICT enabled or delivered programs and services or to facilitate the effective use of the access available through the site, to achieve broader social or economic goals or activities. Thus the telecentre site is not simply a place of "access", rather it is a means through which access is obtained in the context of serving some other sets of organizational, social or other goals and the specific design, staffing, deployment and development of the telecentre has been developed to support this.

From a CI perspective the degree to which the telecentre is developed and structured so as to be truly representative of the broad community and its objectives through the use of ICT, is of considerable importance. Telecentres which are established as for example, to support externally directed services or programs including e-Government services are important to communities. However, having those telecentres as part of on-going community processes where services are identified and then managed through local capacity is much more supportive of long term local development. As well of course, the latter make it much more likely that local resources will be assigned to the centre and thus ensure its long term sustainability.

As well, a key element in the longer term development and effective use of telecentres is their developing networked relationships with other telecentres (and through this with other communities). It is through the development of these distributed self-directed community-based networks that the capacity for on-going self-development and self-management of a range of services (particularly those where local management is a means to achieve optimal program effectiveness and efficiency) can be achieved. In this way for example, economies of scale in services and transactions may be realized.

Universities and colleges play a special role in information societies (or those aspiring to be information societies). They are of course, places where knowledge and information can be found and more particularly where those who possess or can easily and effectively acquire information and knowledge are found in quite high concentrations. And this is particularly true in areas which are somewhat more advanced than is more traditional or stagnant communities or environments.

Also, universities and university people tend to have access to a degree of discretionary resources of particular value from an ICT perspective – computer and, Internet access, software, access to the skills to manage and deploy information technology cost-effectively, and so on. Finally, among the resources available is a degree of discretionary time which can be devoted to voluntary activities or, in the case of students, can be assigned to support community activities. While none of this is inevitable or unique in fact, in many environments universities and colleges may be the only real resource base that communities can access in support of their ICT efforts.

Thus, it is not surprising that many of the pioneer CI initiatives have come from universities or from university faculty looking to make practical and socially beneficial use of their skills and knowledge. As well, it should be added, communities and their use of ICT provides an ideal test bed and experimental environment for certain applications and developments which can in turn lead to broader developments and even commercial products.

## *What is the relationship between CI and ICT for Development?*

**KEYWORDS:**
ICT FOR DEVELOPMENT (ICT4D), APPLICATION, INFORMATION SYSTEMS, SYSTEM DESIGN, LESS DEVELOPED COUNTRY, E-READINESS

At one level CI is one of the strategies and directions available to be used within the broad context and framework of ICT for Development (ICT4D). In that sense ICT4D is the generic application of ICT in support of the development process. However, CI at another level is an implicit critique of the conventional approaches to ICT4D. Most of those and including the agencies and even NGO's concerned with ICT4D begin with knowledge of and skills with ICT which they attempt to find ways of transferring to local communities or to societies in general.

A CI approach however, is one which ideally begins with the local community identifying a need or a possible application and then beginning a process of working with those with the requisite skills to respond to or satisfy that need always within a context where the local community is in control and is directing the process of its own technology enablement. This approach to ICT is of course, one which those involved in Information Systems will recognize as being more or less directly parallel to the overall design and deployment of systems within industrial and work contexts (where of course, rather than bottom up processes, the "client" for the system, responsible for its design and deployment, is corporate management). The overall CI approach here, which is based on a strategy of system design in response to user need and user feedback is one which is well understood in the "systems" context but little understood or implemented in the "development" context.

The utilization of a CI approach to "development" overall (rather than one specifically limited to ICT4D) thus represents a departure from conventional development approaches and practices and, to a degree, presents an implicit critique of these. Rather than CI being only one variety or flavour of ICT4D one can look on CI as a direct substitute or competitor (in the policy or programme sense) with conventional ICT4D. In a jurisdiction which undertook to introduce a CI approach to development and ICT (as has been discussed quite intensively with at least one Less Developed Country) it is quite conceivable that CI would become the overall way in which ICT4D proceeded, and CI itself would become the way in which ICT4D was defined in that context.

In this sense CI has to be seen as an alternative to the e-Readiness approach which has been promoted so actively by the World Bank among others. E-Readiness focuses attention on a limited set of background conditions indicated as being "necessary" for successful ICT implementation. The difficulty with this approach is that it assumes that those conditions are capable of being identified, isolated and responded to. Once appropriate interventions have been made, then it is presumed ICT implementations can be effectively realized. However, this approach fails to recognize the dynamic element in bottom up strategies for ICT appropriation, and that it is precisely this appropriation and those strategies and processes which are the necessary component for ICT to have an impact on local economic and social development. A "deficit" approach to enabling with ICT i.e. one that suggests that there are simple identifiable social deficits to overcome in order to make ICT happen at the local or regional (or national) levels obscures the very real element of social i.e. community processes which are increasingly being recognized as central elements in the achievement of sustainable economic and social development.

## *What is the relationship between CI and Community Development?*

**KEYWORDS:**
COMMUNITY DEVELOPMENT (CD), SELF-ORGANIZATION, EMPOWERMENT, COMMUNITY APPROPRIATION OF ICT

Community Development (CD) is a particular approach to communities which has a specific concern with issues of self-organization, self-management and power. As such CD is to a degree on a parallel track with CI in that a number of the processes which CD might undertake within a community might also be those which are being undertaken in the course of the implementation of a CI project or activity. But in this instance CI and CD are potential allies rather than competitors. The goals of CI and CD relative to communities are the same i.e. self-development, self-management, and empowerment. Thus CI (or CD) could build on the skills and strategies of CD (or CI) and where available align itself to CD (or CI) activities or equally to provide additional resources and skills to CD (or CI) processes.

Equally, CI initiatives benefit significantly from having access to CD processes and CD skills within communities and CI applications/implementations can be most effective (and least resource intensive) in their implementation where they can build on pre-existing CD activities. In that sense then CI and CD are complementary and where feasible, syn-ergistic; with CI being necessarily more specific in its focus on ICT as a single skill and resource structure while for CD, CI is but one among a range of skills and potentially available resources.

However, in many if not most instances CI processes run parallel to CD processes with different memberships and leaderships. The reasons for this are complex and vary from community to community, but in many instances the skills and experiences (and orientations towards technology) differ quite dramatically at the local level between those who are ICT oriented and those who may have a more traditional approach to enabling and empowering communities. It is not that one is better or worse than the other but just that some are convinced that face to face processes are a necessary precondition for enabling communities and that self-management with ICT is only one among a possible range of instruments in support of this. Those with a more technology focus might say that there are elements of being enabled and empowered through ICT and particularly in relation to more advanced activities and services (or when dealing with more developed systems) that can only be accomplished through the use of ICT.[33] A third approach might be termed that of community *appropriation* of ICT where the community has moved beyond simply seeing ICT as one among a range of tools and has become sufficiently knowledgeable and comfortable with the technology that it is able to work effectively both in face to face and technology enabled modes depending on the circumstance and the desired output. In this latter instance, the community itself is beginning to see novel ways in which ICT can be used as an on-going element in its self-development process.

## *What is the connection between Community Informatics and Building Alternative Structures of Production?*

#### KEYWORDS:
INNOVATION SYSTEM, COMMUNITY INNOVATION SYSTEM (CIS), EFFECTIVE USE, SOCIAL NETWORKING, FLEXIBLE NETWORKS, NETWORKED ORGANIZATIONS, ONLINE NETWORKS, DISTRIBUTED PRODUCTION

At the community or local level there is the opportunity to "innovate", if only in the form of developing new (for the area) types of businesses, production processes, and markets. Similarly to other forms of "innovation systems",[34] the Community Innovation System (CIS) requires access to advanced levels of information and skilled knowledge workers for assimilating and implementing the knowledge being identified. In the community context of course, the scale and level of the information being assimilated is of a more modest nature than for regional or national systems.

An important element of a CIS in addition to the knowledge from which the innovation springs, is the capacity of the local productive and cultural system to absorb and make effective use of the information (and the technology which is often a component of this) which is being made available. In the community context, this capacity is closely linked into local cultural practices and norms. Many communities, particularly smaller and more isolated ones are often characterized by an unwillingness to experiment or to absorb new information or techniques. In addition, many communities and particularly those without a tradition of knowledge based industries or considerable numbers of locally based knowledge workers may be suspicious of new information and indifferent to themselves or their children obtaining the education and particularly advanced education from which innovation can spring.[35]

This approach to innovation is often based on processes of social networking which are to be found within local and community contexts where such informal associations and networks provide the very substance of connection between individuals. While at the community level such networks are primarily the basis for social inclusion and adhesion, there is little except inertia and tradition which prevents these from becoming a basis for local innovation and the foundation for community innovation systems. In fact, it is precisely these types of connections which have provided the platform and mode of operation for some of the most economically successful and innovative of local communities through the integrated use of local social networks as networks for managing local production and distribution in what is generally referred to as "flexible networks".[36]

Technology as an exogenous, i.e. external, factor which enters into the community may in many cases be a support for innovation (in other cases it may be a factor in community disruption as well). The opportunity with a Community Informatics[37] approach is for the community to have some direction and responsibility i.e. "ownership" of the innovation and the innovation strategy. The use of a CI technology strategy ensures that "innovation" is done by, with and in the community and not simply something that is done "to" or "for" the community. By adopting a CI approach, there is a degree of assurance that the process of innovation will become an on-going element of community life and activity rather than a once for all investment in, for example, a single high profile "innovating institution".[38]

New types of networked organizations may be created. These could be structured as hubs and multiple self-sufficient nodes. Collaborative specialization, information dispersal and multiple or distributed ownership, decentralized and horizontal support structures, and a high degree of local self-suf-ficiency (and thus structural redundancy/survivability) characterizes these new organizations. These structures also allow for a speed of adaptation, highly efficient (low friction) horizontal rather than vertical information flow, and the economies of mutual rather than functional support. Client needs can be responded to more immediately, both geographically and culturally, creating powerful and globally competitive marketing opportunities.

A *flexible network* is a group of two or more firms which have banded together to carry out some new business activity that the members of the network could not pursue independently. "Flexible networks" as well, gain advantage from geographic or cultural distinctiveness and from being a component of a larger network of producers, even when the linkages are largely "virtual". The network can involve similar firms which band together to share the costs of developing a new product or market, or dissimilar but complementary firms which collectively approach the capability of a vertically integrated large firm. Typically the nature of the cooperation within the network is carefully defined so as to preserve each firm's independence and original lines of business. The duration of the collaboration may be very short (or very long) and limited to a particular project for a single customer with a new network being assembled with the best configuration to meet the needs of the next customer as might be required.

Among other strategies, ICT supports the formation of online networks and networking for distributed production and economic development. Technology for example, allows for continuous communication; work sharing; remote administration and management; and, seamless presentation and marketing of multiple centres as a single entity to the world. As well, this could include the electronic (and remote) coordinating of production, in turn allowing for an optimizing of the selective advantages within a network and using the larger scale capaci-ties of the network to undertake more elaborate activities. This approach could be a major opportunity for local economies that previously had been limited by their access to specialized skills and their small and dispersed populations.

A flexible production network is not just a joint venture among several firms. The nature of the collaboration tends to be deeper in a true network, and one form of collaborative endeavour tends to lead to others. Shared input procurement to get large scale cost breaks may lead to joint bids or a common work force training program. These new organizations would utilize the strengths and competitive advantages of existing economies and resources in local areas. In turn, these firms would be highly adaptive to external economic conditions yet assist the development of local economies. On a local or regional scale, there would be an increasing use of information tied to sophisticated market demands, leading to an increasing need (and opportunity) for complex elaborations of products and services, and the capacity to integrate clients directly into the supply chains of dispersed producers.

This in turn would map onto the strengths and competitive advantages of existing local enterprise efforts. Highly adaptive responses to external economic conditions would help the local economy to evolve towards information intensity, increasing complexity and functional elaboration while integrating clients directly into dispersed supplier chains. The resulting disintermediation between user and supplier is precisely what many are predicting as being the organizational model of the marketplace of the immediate future. Several of these networks have been created and have succeeded in creating synergies and economies of scale based on networked coordination rather than organizational structuring; distributed divisions of skills, responsibilities and efforts within the network; and effective distributed marketing, quality control and, even research and development efforts.

## *What is the relationship between Community Informatics and Open Source?*

**KEYWORDS:**
OPEN SOURCE, COMMUNITIES OF PRODUCTION, ELECTRONIC INFRASTRUC-
TURE, CODE, NETWORKED COMMUNITIES

There is a real opportunity for CI to make common cause with those supporting Open Source and Open Access research since so many of the elements, both of communities networked in the manner in which CI understands these and through the dialectical nature of technologically enabled community action are similar to or inclusive of Open Source research and practice.

There are direct parallels between ICT enabled "communities of resistance" and the "communities of production" within which Open Source software is being produced. Parallel to the creation of a totalizing (monopolistic) retail structure of distribution that Wal-Mart has built on its centrally-controlled electronic infrastructure, is the monopolistic and totalizing infrastructure-production control systems presented by Microsoft through its Windows operating system.

Microsoft (MS), in a way similar to Wal-Mart uses internal networks for production and, as a progenitor of globalized IT-enabled production and networking, provides *closed* i.e. centrally structured and controlled access software code whose use and application is widely, even universally distributed (and enforced) through Microsoft's monopolistic control of Windows. Even though the individual uses (instantiations) of the application are highly individualized (contextualized), nevertheless the *code* is highly controlled by Microsoft and, through the invisible networks of Personal Computer (OEM) producers and their relationships with other equipment suppliers, MS is able to maintain and reproduce this centralized control for its own monopolistic gain.[39]

The opposition (resistance) to MS and (pre-Google) the only effective competitive force that has arisen is the Linux/Open Source community of non-centrally-controlled i.e. dispersed, electronically enabled "communities" of software code writers who have created a framework of production and of (more or fewer) products "competitive" to the Microsoft software suites and offerings. In some sense these "communities" present a clearly identifiable alternative model of electronically enabled (networked) "communities". Most notably, rather than being centrally controlled with top down decision making and structured around contractual linkages these communities are founded on relationships of voluntary participation, peer to peer engagement and operate within a distributed (and to a degree) consensus based organizational form.[40]

The structure here, rather than being one of "networked individualism" is in fact one of "networked communities" which quite evidently have the resilience, breadth, depth, innovative capacity and persistence to represent a very significant competitive threat to perhaps the world's most powerful Information Age corporation.

## *Is there an "Urban" Community Informatics?*

**KEYWORDS:**
COMMUNITY PROCESSES, URBAN AREAS, COMMUNITY, GLOBAL CITY TOP LEVEL DOMAIN NAMES, ELECTRONICALLY ENABLED "COMMUNITIES", ICT ENABLEMENT, RURAL AREAS, ONLINE COMMUNITIES

Perhaps not surprisingly CI developed initially in response to the attempts to find ways of bringing the benefits of ICT to those living in rural and remote locations. The link between ICT and community processes and particularly the identifica-tion and alignment of ICT activities with community dynamics is of course, more easily undertaken in rural or smaller communities where traditional "community" processes are more readily identifiable.

This is not to say however, that there is no role for CI in urban areas although at a conceptual and analytic level the task may be more complex.[41] The challenge in an urban environment is to identify "the community". In some instances neighborhoods may function as communities.[42] In some other contexts the entire urban environment may function as a "community". The work being done to create "global city Top level domain names" (GC-TLD's)[43] is an example of this as is the Milan Community Network's (RCM's) political dialogue forums.[44] In still other and perhaps more common instances, it is the virtual or electronically enabled "communities" developed within a specific urban environment which may function as the community agent for a CI application.

Another element of urban CI is that in an urban environment an individual is most likely to be a member of multiple communities, some of which are electronically enabled but others of which aren't. Equally in this environment, the individual can be expected to be participating in and have loyalty (of varying degrees of course) to these multiple communities. What this means in the urban context is that the overlap between the social processes of communities and the process of ICT enablement is going to be of a rather more hit and miss fashion and far more likely to be selective and user driven than in a rural environment where there is a much greater overlap and integration of the multiple communities/multiple loyalties of which an individual or a family might be part. Thus in the rural environment the possible overall impact of ICT in and through the community processes could be much more intense and transformative than in the urban environment although in the urban setting the larger numbers of par-ticipants may result in a larger total number of individuals very actively engaged in specific communities than could be found in rural communities.

Also, if one believes (as I do in part) that there is a more or less general or universal desire for participation in community (with the resulting relationships of (limited) intimacy and trust that comes with this) then the more intense and integrated (and overlapping) community connections in rural areas will mean that those in urban areas who seek such a connection (and the values of these communities) may in fact look with even greater hunger to the online communities as a way of getting their community "connectedness". Those in rural areas are able to satisfy these desires in other (face to face) ways.

## *How does Community Informatics Contribute to Social Capital and what about "Sustainability"?*

**KEYWORDS:**
SOCIAL CAPITAL, COMMUNITY INTERACTION, ICT APPLICATIONS, SUSTAINABILITY, ECO-SYSTEM, CI APPLICATION, SERVICE DELIVERY, COMMUNITY BASED SUPPORT INFRASTRUCTURE

CI both contributes to and benefits from social capital. If we understand social capital to be the result of community interaction and community connections, then CI as a means for enhancing and extending community interaction (both within and between communities) and allowing for the elaboration and proliferation of community connections again both within and among communities, can be seen as a significant contributor to social capital. As well of course, social capital as understood by the volume of these connections allows for or facilitates the deployment of ICT applications and can contribute to the overall success of these applications by eliciting contributions of time, skill and resources from community members for a specific deployment as for example in training, management, maintenance and so on.

The use of these connections and this connection for training and skill development, for maintaining and extending networks, for identifying and facilitating access to skills are equally contributors to community social capital. The sometimes subtle interplay between CI and social capital formation has been the subject of some of the most interesting research work done in CI as for example that by Fiorella de Cindio and her colleagues in Italy (and the Milan Community Network) who looked at the role that communities can play in design innovation in the area of e-Government and the role that ICT can play in enhancing democratic participation within communities;[45] and Lyn Simpson in Australia seeing how even the simplest of ICT can make a significant contribution to the development of social (and human) capital as evidenced among the participants in a virtual network/community in rural Queensland.[46]

The question whether CI can be sustainable is really the question of whether CI applications and a CI approach can be sustainable. The question in fact might be turned around – is it likely or even possible for a non-CI application; that is can an application that is not firmly rooted in community processes and building on community commitment and contributions be sustainable in the longer term (without subsidy). The very nature of a CI application is that it is built of and from community identified needs, responding to and incorporating community resources and skills and designed to be part of local community economic and social "eco-systems", including the broad range of suppliers, resource providers, markets, users and others.

It is through being firmly involved in, contributing to and consuming as part of a local eco-system, that a CI application

(as with other community based initiatives) in fact can be sustainable i.e. being able to survive in the longer term based only on local resources – including voluntary and in-kind contributions as well as fee for service payments. If a CI application is in fact, part of such a local eco-system and is responding to a real felt need at the local level, then the likelihood of the application surviving over the longer term i.e. being sustainable is most certainly the case. Since sustainability is primarily an issue of social i.e. community support, it is more likely for a CI application to be sustainable than a top down and externally maintained/supported/funded application.

It might also be noted here that a CI application is in many service areas the only delivery approach that could be sustainable. The notion that external resources would be available in sufficient volume and over the longer term to, for example, provide primary health care to all in the world through the use of professional staff paid at urban/metropolitan salary levels and with an urban and developed sector support infrastructure, is almost certainly pie in the sky. The design of service delivery and support systems enabling the use of local resources and local infrastructures (with commensurately lower cost levels), integrated with electronic information delivery and management and service support systems, could allow for service levels at least at the primary level to be maintained with existing and even local resources.

Thus for example, the cost of totally professionalized and hospital based health care continues to increase exponentially in developed countries (well beyond the levels of overall GDP increase in most instances) and as the demand for comparable service levels continues to extend into broader segments of the population throughout the world the situation with respect specifically to health service, is almost certainly not-cost sustainable even over the medium term. The re-development of health services built around a community based delivery and support infrastructure with ICT as a central component for supporting information delivery, training, diagnostics and counseling is almost certainly the direction which must be taken.

*Is there a Wireless Community Informatics?*

**KEYWORDS:**
WIRELESS COMMUNITIES, WIRELESS COMMUNITY INFORMATICS, COMMUNITY NETWORKING, COMMUNITY CONNECTIONS, INTERNET SERVICE PROVIDER (ISP), HOTSPOT, ILE SANS FILS (ISF)

Issues around wireless communities as communities and whether there is a wireless Community Informatics arise partially from the peculiar artifact that many of the early community wireless innovators, while having somewhat parallel backgrounds to the early community networking innovators (progressive politics, university education, technical proficiency, youthful) and even often referring to themselves as "community networkers", yet had little knowledge of (or interest in) more traditional or long standing community networking advocates or institutions.[47]

The difficulty with thinking about a "wireless community informatics" is that wireless as an infrastructure is necessarily virtual and placeless thus immediately making the creation of community connections (the normative integration necessary for community formation) exceedingly difficult. Wireless users are simply those who use a wireless connection to obtain Internet access and need have no other links or connections to other Internet (or Internet Service Provider (ISP)) users. However, the fact that a wireless connection does have some degree of geographical anchoring (individuals gain access through a geographically anchored "hotspot") opens up the possibility of creating a degree of interconnection among users over and above that of a random and anonymous (to each other) group of users of a single ISP.

Certain early ISPs and subsequently AOL, among others, identified a similar difficulty and a similar opportunity and used their role as ISPs to interpose a screen between their users and the Internet. In this way they forced their users to interact with this screen as part of their virtual activities. In most cases they gave that screen a content (generally advertising). In this way the possibility of a many-to-many interaction among those being forced to interact with the interposed screen was created. In the context of wireless, these interposed screens have the possibility of being created and interacted with at the level of the "hotspot". This has given, the ISP (or hotspot operator) an opportunity to develop a degree of inter-connection among these wireless users and to facilitate the formation of a degree of "community" interaction among them or, as in the case of AOL (or Ile Sans Fils – ISF – in Montreal) the means to use this limited amount of interaction as a way of, in one case (AOL), presenting advertising to users (as they are accessing the interposed information) or as is the case of ISF, using this interposition for local promotion and development.

The degree to which this interposition and hotspots or interconnection between hotspots can be used as the basis for social mobilization or community service delivery is the degree to which one can begin to conceptualize and formulate a Wireless Community Informatics.

*What are the Challenges and Opportunities facing Community Informatics?*

**KEYWORDS:**
BOTTOM UP DEVELOPMENT, ICT DEPLOYMENT, INFORMATION SOCIETY, INFORMATION INTENSIVE SERVICES, MANAGEMENT INFORMATION SYSTEMS, PRACTICE, INSTITUTIONALIZATION, CI AS A MOVEMENT

CI has the opportunity to provide direction and a basis for rethinking the ways in which communities engage with technology, the direction for the developmental implementation of information systems, and perhaps most important to provide leadership in re-affirming the role and significance of communities in a modern technologically enabled society. At its most fundamental CI presents a critique of conventional approaches to development (a shift from top down to bottom up in the way in which ICT is deployed in society i.e. to empower collectivities as well as individuals) and suggests a new direction for the democratization of the information society (using ICT to transfer responsibility and authority to communities and away from central institutions).

It also represents what will in the medium and longer term prove to be the only feasible (sustainable and cost-effective) way in which to organize and widely distribute information intensive services. That being said the opportunities for CI both as a practice and as an academic field for research and development are outstanding. My own feeling is that CI could (and should) develop as a parallel stream to Management Information Systems within "Information", "Information Studies" or "Information Systems" programs in universities and colleges; however with a stronger element of a "practice" component which in this instance might be cross connected with say Social or Public Administration programs, IT design and development programs and service design and development programs.

However, there are also a variety of challenges facing Community Informatics. Perhaps the greatest immediate challenge is the problems associated with institutionalization. CI started out as an "outsider" phenomenon… gathering together academic outsiders and engaged techie practitioners but particularly academic outsiders who were dissatisfied with the specific academic disciplines in which they found themselves and especially because those disciplines seemed not to be taking into account the opportunities and risks associated with the new ICT notably as they might impact on marginalized and developing world populations. Contrary however, to most outsider academics, CI researchers are not necessarily of a "critical" persuasion; that is, while they might be critical of the particular academic framework in which they found themselves either from an institutional or from a conceptual perspective many were intrigued and challenged by the opportunities for broader social change which they saw ICT as representing. Thus, rather than being pessimistic about the changes being undergone they saw many of those changes as being (at least potentially) for the better and they saw their own work as helping or intervening so as to enable or nudge these developments in desirable directions.

I think it was Randall Pinkett at a meeting in Colorado Springs that first suggested to me that CI was not a discipline or a practice but rather a "movement" and I've been intrigued by that idea ever since and have noticed that others have also begun to use that terminology. I think that CI has some of the characteristics of a movement in that there tends to be a degree of ideological adherence and struggles concerning consistency and even doctrine within the CI community. More important than that however there seems to be at the core of CI a vision of how the world might look (with communities as enabled by technology and the restructuring of social and political power that would be associated with that), and some sense of how and under what circumstances that might be achieved – through bottom up approaches to technology implementation and broad social appropriation. As well, there is a sense that a CI perspective is a moral (or even political) perspective rather than simply a set of concepts, models and techniques and, as a moral direction, there is a sense that involvement with CI is more than simply accepting an academic direction, it is also buying into a directed and collaborative quest.

*What is likely to be the long term impact of Community Informatics and Why Does It Matter?*

**KEYWORDS:**
COMMUNITY INFORMATICS, TRANS-DISCIPLINARY, PROBLEM FOCUSED, PRACTICE, SOCIAL AND ECONOMIC DEVELOPMENT, COMMUNITY EMPOWERMENT, INCLUSION, ELECTRONICALLY ENABLED SERVICES, E-HEALTH, INFORMATION SOCIETY, SOCIAL NETWORKING SOFTWARE, BOTTOM UP IMPLEMENTATION

Community Informatics is one of a number of emerging trans-disciplinary approaches which for the most part are "problem" or output focused. A lot of these are technology related – gaming is another example, but CI is perhaps the most ambitious since it is trans-disciplinary at a level which incorporates the social as well as the behavioural rather than simply the behavioural.

Certainly the future for CI as a practice appears to be a bright one, with accelerating interest and an increasing number of adherents both as researchers and as practitioners. Providing technical resources and strategies designed for and implemented by local communities, is such an obvious approach and so obviously superior to top-down approaches for achieving social and economic development goals at the local level that it is difficult to see how this would not become the norm in at least socially progressive developing country contexts.

The issue of community empowerment with ICT is of course a significant element of the broad penetration and use by local communities of ICT and this will not always be seen in a favourable light by existing authorities. How this will be handled remains to be seen (it is one aspect of the broader transformation of political institutions and processes in response to computerization) but as the need for broader inclusion into the social and economic fabric becomes ever more evident and the link in this with ICT is established, the trend towards CI as a basic policy framework for ICT deployment and use would seem to be very likely.

Whether we will see a broader transformation in the delivery of electronically enabled services into communities within a CI framework is rather less obvious. The transformation that would be required is one that affects the service system at all levels and all stages and to a degree requires the re-thinking of the nature of the service itself including the professional structures which support it. If we think of education as an example (it could be health or the range of other generally publicly provided services) then the notion of shifting the balance to community based learning, utilizing community knowledge resources and personnel as the first line of service providers is an obvious one. In this instance, those up the bureaucratic structure would necessarily need to shift from monitoring and control of professionals to enabling community processes and where professionals, rather than being direct service providers would become supporters of local processes of service delivery at least at the more basic level of skill and requirement.

The area of health however, may be a special case since the cost of health care is becoming increasingly unsupportable and the need to find lower cost alternatives to at least the basic level of service provision is becoming ever more evident. The provision of health care through community based facilities and service providers supported through specially designed content and software would appear to be almost inevitable particularly and initially in least developed countries. It is likely in more developed ones as well as the message comes through that the aggregate level of health care achieved through this bottom up approach can be dramatically enhanced in this way. The difficulty of course, comes from entrenched positions and interests on the part of current professional and bureaucratic structures and these are the most resistant to change. Any overall change will thus need to be done in large part through engaging and overcoming resistance from these quarters.

The question of if or why Community Informatics "matters" and to whom goes to the very core of the nature of the Information Society and of the options that are available for its development. That there is a desire to use the tools of ICT to maintain, enhance and extend communities is I think, unarguable given the success of social networking software which to a degree is based on precisely these objectives. Equally, there is an emerging broad recognition that only through a concern for bottom up implementation and effective use of ICT can these become useful tools as part of the process of economic and social development.

CI matters because there is a need not only to do but also to systematize and to understand what is being done. In the absence of this understanding then the processes of achieving success are at best random and at worst may result in a continuous wastage of resources, time and credibility as mistakes are never learned from and so repeated, and as successes are never captured and built upon to realize further horizons.

# Resources

There are a range of resources available concerning Community Informatics. Those with a continuing interest might begin with the Wikipedia site *http://en.wikipedia.org/wiki/Community_informatics* and the Community Informatics Research Network (CIRN) website *http://www.ciresearch.net/*

The Journal of Community Informatics *http://ci-journal.net* is regularly publishing a range of peer reviewed research articles in the area.

Community Informatics currently functions as a loose network focused around the electronic discussion lists, *Community Informatics@vcn.bc.ca* and *CIResearchers@vcn.bc.ca*. To subscribe send a message to: *sympa@vancouvercommunity.net* with a blank subject line and in the body write: subscribe communityinformatics and/or ciresearchers. Archives are available.

# Notes

1 The title is a parallel to Rob Kling's famous article presenting the case for a "Social Informatics" cf. R. Kling, "What is Social Informatics and Why Does it Matter?" *D-Lib Magazine*, January 1999, Volume 5 Number 1.

2 M. Gurstein, Community Informatics: Current Status and Future Prospects – Some Thoughts (source: adapted from *Community Technology Review* winter/spring 2002).

3 For a fuller discussion of this see below.

4 Cf. also M. Gurstein (Ed.) *Community Informatics: Enabling Communities with Information and Communications Technologies*, Idea Group, 2000; L. Keeble & B, Loader (Eds.) *Community Informatics: Shaping Computer-Mediated Social Networks*, Routledge, 2001; and the Proceedings of the IT in Regional Areas Conference: *Using Informatics to Transform Regions* (Eds. S. Marshall, Wal Taylor and Xinghou Yu), (2004).

5 Cf. M. Castells, *Rise of the Network Society: The Information Age: Economy, Society and Culture* – 1996 – Blackwell Publishers, Inc. Cambridge, MA, USA.

6 Much of this is drawn from a previously unpublished working paper presented to the CRACIN Workshop Toronto, 2006.

7 B. Wellman, A. Quan-Haase, J. Boase, W. Chen, K. "The Social Affordances of the Internet for Networked Individualism", *Journal of Computer Mediated Communication*, JCMC 8 (3) April 2003.

8 Van Dijk presents a very useful explication of Wellman's theory as follows: This means that the individual in one of its roles increasingly is the most important node in the network and not a particular place, group or organization. The social and cultural process of individualization is strongly supported by the rise of social and media networks. Using them the individual creates a very mobile lifestyle and a crisscross of geographically dispersed relations. Every mobile phone user knows that (s)he does not any longer reach a place, but a particular person in one of its roles. This practice may be very liberating and self-empowering, but there also is a less positive side to it. Less and less people have a view of us as a whole person: one only knows one or a few sides of our personality playing a particular role (Wellman, 2000). Presently, the last refuge where one is supposed to know each other more completely, the family household, is dispersed also. In families husbands and wives, parents and children are engaged with ever more different activities in other social and media networks. Effects to be observed might be an increase of loneliness, alienation, uncertainty and the feeling of not being understood by others. This might happen in spite of, or because of (?) the virtual explosion of means of communication available.
http://www.gw.utwente.nl/vandijk/research/network_theory/network_theory_plaatje/a_theory_outline_outline_of_a.doc/

9 In this as in other areas, when we are discussing externally-driven networks based on centralized decision making, we should include as direct parallels the processes of the transfer into electronic format of Government services (e-Government) without the parallel development of enhanced means for enabling citizen participation and control at the community level of these services (e-Governance). For a further elaboration of this discussion see M. Gurstein, D. Schauder, and W. Taylor "E-Governance and E-Government", for *In-* *ternational Conference on Engaging Communities*, Brisbane, Australia July 2005.

10 B. Wellman, K. Hampton, "*Living Networked in a Wired World*" Contemporary Sociology, 1999 – chass.utoronto.ca

11 http://pages.ebay.com/help/policies/hub.html?ssPageName=home:f:f:US

12 Cf. The extremely interesting discussions which are emerging concerning the notion of "code as law" for example those by Lawrence Lessig in his electronically published book "*Code and other Laws of Cyberspace*" http://www.code-is-law.org

13 Cf. eBay op.cit. A person suspended from eBay loses all membership privileges. A suspended individual is not permitted to participate on the eBay site using any existing account, or to register new accounts with eBay. A suspension from eBay may be for a fixed length of time, indefinite, or permanent. Suspensions remain in effect until removed by eBay. http://pages.ebay.com/help/policies/rfe-previously-suspended html
It has been indicated that currently some 250,000 individuals and businesses are now deriving a majority of their livelihood from transactions on eBay. Thus being *suspended* from eBay as described above, without right of notice or appeal is potentially an extremely significant sanction and gives those enforcing such rules enormous economic and social power.

14 Cf. *Journal of Community Informatics*, Special Issue on CI and Systems Design.

15 Cf. M. Gurstein & T. Horan. "Why Community Information Systems Are Important to the Future of Management Information Systems and The Field of Information Science (IS)" *The Gordon Davis Series on the Future of Information Systems Academic Discipline: Opportunities and Directions*, 2005.

16 Cf. Lyn E Simpson, "Community Informatics and Sustainability: Why Social Capital Matters" *Journal of Community Informatics* Volume 1, No. 2, 2005.

17 A number of companies in the DotCom period and immediately after created on-line forums giving customers the opportunity to present feedback to the company and with the intention of creating "communities" around the various products or brands as is promoted by Hagel and Armstrong in their very influential book, Net Gain: Expanding Business Through Virtual Community. Most of these were quickly shut down when the customers began to interact with each other to form groups of customers many of which were directly critical of individual company offerings. A number of these eventually re-emerged in the "xxxsucks.com" phenomenon as in:
http://www.mycarsucks.com/ for example.

18 These processes have been quite well examined as for example M. Gurstein, "*Community Informatics: Enabling Communities With Information and Communications Technologies*" and the variety of articles in the *Journal of Community Informatics* http://ci-journal.net

19 "Globalization" as a term has no standard and universally recognized definition. Rather it can be understood to occupy a general conceptual space and is adapted to particular circumstances as required. A fairly typical definition as applied within the context of Information Systems would be the following: "globalization of business refers to a qualitative departure from traditional approaches to doing business internationally. An important distinction is the size of the new business entities. Another, and significantly more interesting aspect is the attempt to set up such entities in various countries to functioning as single, "seamless" business operations. For example, while a corporation's market in international trade is usually considered to be composed of many, country-defined markets, in the globalized approach it is defined as one, huge globe-encompassing mammoth. Closely related to this notion is the global business corporation's approach to management of business operations in various countries, as elements of a unified system, regardless of the location and national boundaries. A significant implication of this approach is the expected ease of transfer of goods, services, capital and labor across the globe, unencumbered by excessive local and national regulations. http://aabss.org/journal2002/Mahdavi.htm

A further and more Information Systems and less academic definition would be the following: "We define globalization as the responsible development of a geographically balanced network of business units that are fully integrated within both our worldwide business structure and within the local societies in which they operate". http://www.st.com/stonline/company/annual/fy01/p07a.htm

20 This "control" takes the form of either contracting or not contracting i.e. either a company conforms to the technical requirements and standards of Wal-Mart or it doesn't do business with Wal-Mart and, given the massive significance of Wal-Mart as a purchaser this means that conformity is not voluntary but a compulsory aspect of staying in business. http://www.ladlass.com/ice/archives/007533.html

21 Much of the conceptual insight into the role and operations of Wal-Mart and other electronically enabled enterprises comes from the very useful introductory to e-Business by Kalakota and Robinson. "*e-Business: Roadmap for Success*", Addison and Wesley, 2002.

22 The integration of the sales "chain" with the supply chain is probably Wal-Mart's most significant single innovation. Developing the capacity to directly link sales with supply has allowed it to achieve massive efficiencies by effectively eliminating the need for inventory and warehousing. The "joke" in the industry is as follows: Q. Where is Wal-Mart's warehouses? A. The US Interstate highway system.

23 http://walmartwatch.com/

24 Cf. Wellman's references to contractual or "gemeinschaft" relationships as per Durkheim's notions as the defining characteristic of his "networked individualism" postulate. Wellman and Hampton, op.cit.

25 Gurstein and Horan op.cit.

26 http://rkcsi.indiana.edu/

27 Adapted from M. Gurstein "Effective Use: A Community Informatics Strategy Beyond the Digital Divide", *First Monday*, December 2003. http://www.firstmonday.dk/issues/issue8_12/gurstein/index.html

28 http://searchsmb.techtarget.com/sDefinition/0,,sid44_gci214062,00.html

29 Canadian International Development Agency, *CIDA's Strategy on Knowledge for Development through Information and Communication Technologies (ICT),* http://www.acdi-cida.gc.ca/ict

30 Cf. Fiser, Adam, Clement, Andrew, & Walmark, Brian. (2006). *The K-Net Development Process: A Model for First Nations Broadband Community Networks*, CRACIN Working Paper No.12.

31 *B. L. Bell, P. Budka, & A. Fiser* "*We Were On the Outside Looking In*" *MyKnet.org: A First Nations Online Social Network in Northern Ontario, 5th CRACIN Workshop Concordia University, Montreal, June 20-22, 2007.*

32 Association for Community Networking, website: http://www.afcn.org/taxonomy_menu/1/10?page=13

33 Cf. R. Stoecker, Is Community Informatics Good for Communities? Questions Confronting an Emerging Field. *Journal of Community Informatics* Vol 1, No 3 (2005).

34 For a more complete discussion of Innovation systems in this context see M. Gurstein, "A Community Innovation System: Research and Development in a Remote and Rural Community", in D. Wolfe and A. Holbrook (Eds.), *Knowl-* *edge, Clusters and Regional Innovation Systems*, Queen's-McGill University Press, 2002**.**

35 For a more extensive discussion of these points see M. Gurstein (2002) op.cit.

36 Cf. M. Gurstein, http://www.firstmonday.dk/issues/issue 4_2/gurstein/ and M. Gurstein, "Community Informatics, Community Networks and Strategies for Flexible Networking" in *Community Informatics: A Social Agenda for Technology*, Brian Loader and Leigh Keeble (Eds.) Routledge, London, 2001.

37 Cf. M. Gurstein, Community Informatics: Current Status and Future Prospects – Some Thoughts, Community Technology Review, Winter-Spring 2002.
http://ctr.dyndns.org/article.php?article_id=56

38 M. Gurstein, "Effective Use: A Community Informatics Strategy Beyond the Digital Divide", *First Monday*, December 2003. http://www.firstmonday.dk/issues/issue8_12/gurstein/index.html

39 This of course, is the basis of the recent finding by both the US government and the European Union that Microsoft is guilty of monopolistic practices.
Cf. http://europa.eu.int/rapid/pressReleasesAction.do?reference=IP/04/382&format=HTML&aged=1&language=EN&guiLanguage=en

40 Cf. L. Torvalds – Communications of the ACM, 1999 – portal.acm.org *The Linux edge*, Volume 42, Number 4 (1999), Pages 38-39.

41 The Canadian Community Access Program was one of the first large scale CI programs in the world and not surprisingly the overall objectives of this program were to facilitate Internet access in the vast rural and remote areas (and widely among rural and remote populations) in the country. The CA Program had a striking success in the rural areas – having very widespread uptake, a high measure of voluntary participation and enthusiasm and a significant overall level of participant satisfaction as manifest in political support for the program from local politicians. Based on this experience the Government of Canada determined that it would carry forward with a parallel program in urban Canada and assigned resources including staff and start up funds for this purpose. What this program didn't do however, was to adapt the strategies which had been successful in a rural environment to the urban environment and nor did the Program undertake a clear assessment or develop a clear plan of how it would proceed in this new context. The result was that the program had very very great difficulty getting launched. For reasons of lack of success in finding local sponsors and counterparts (and champions on the ground in the urban areas) the Program itself was almost two years delayed in being launched and once launched had great difficulty in finding "communities" at the local level that could sponsor and sustain these access sites in a manner that had proven so effective in the rural areas. In the urban environment "communities", and thus community structures and processes look and operate quite differently and thus a program that has been successful in rural Canada could not necessarily be expected to be successful in urban Canada.

42 Steve Clift's work with Neighborhood based political action in Minnesota is a good example of this cf. http://www.publicus.net/

43 T. Lowenhaupt & M. Gurstein *Towards City-TLDs in the Public Interest - A White Paper:* http://www.openplans.org/projects/campaign-for.nyc/towards-city-tlds-in-the-public-interest

44 http:// www.retecivica.milano.it/

45 F. De Cindio, L. A. Ripamonti, & C. Peraboni, "Community Networks as lead users in online public services design", *Journal of Community Informatics*, Vol. 3 No. 1, (2007).

46 L. E. Simpson "Community Informatics and Sustainability: Why Social Capital Matters", *Journal of Community Informatics*, Vol 1, No 2 (2005).

47 Cf. http:// www.seattlewireless.net/

# LIST OF KEYWORDS

(Some keywords appear duplicated in order to facilitate the search)

Creative Commons Legal Code
Attribution-NonCommercial 3.0 Unported

    not be considered an Adaptation (as defined above) for the purposes of this License.

c) "**Distribute**" means to make available to the public the original and copies of the Work or Adaptation, as appropriate, through sale or other transfer of ownership.

d) "**Licensor**" means the individual, individuals, entity or entities that offer(s) the Work under the terms of this License.

e) "**Original Author**" means, in the case of a literary or artistic work, the individual, individuals, entity or entities who created the Work or if no individual or entity can be identified, the publisher; and in addition (i) in the case of a performance the actors, singers, musicians, dancers, and other persons who act, sing, deliver, declaim, play in, interpret or otherwise perform literary or artistic works or expressions of folklore; (ii) in the case of a phonogram the producer being the person or legal entity who first fixes the sounds of a performance or other sounds; and, (iii) in the case of broadcasts, the organization that transmits the broadcast.

f) "**Work**" means the literary and/or artistic work offered under the terms of this License including without limitation any production in the literary, scientific and artistic domain, whatever may be the mode or form of its expression including digital form, such as a book, pamphlet and other writing; a lecture, address, sermon or other work of the same nature; a dramatic or dramatico-musical work; a choreographic work or entertainment in dumb show; a musical composition with or without words; a cinematographic work to which are assimilated works expressed by a process analogous to cinematography; a work of drawing, painting, architecture, sculpture, engraving or lithography; a photographic work to which are assimilated works expressed by a process analogous to photography; a work of applied art; an illustration, map, plan, sketch or three-dimensional work relative to geography, topography, architecture or science; a performance; a broadcast; a phonogram; a compilation of data to the extent it is protected as a copyrightable work; or a work performed by a variety or circus performer to the extent it is not otherwise considered a literary or artistic work.

g) "**You**" means an individual or entity exercising rights under this License who has not previously violated the terms of this License with respect to the Work, or who has received express permission from the Licensor to exercise rights under this License despite a previous violation.

h) "**Publicly Perform**" means to perform public recitations of the Work and to communicate to the public those public recitations, by any means or process, including by wire or wireless means or public digital performances; to make available to the public Works in such a way that members of the public may access these Works from a place and at a place individually chosen by them; to perform the Work to the public by any means or process and the communication to the public of the performances of the Work, including by public digital performance; to broadcast and rebroadcast the Work by any means including signs, sounds or images.

- i) "**Reproduce**" means to make copies of the Work by any means including without limitation by sound or visual recordings and the right of fixation and reproducing fixations of the Work, including storage of a protected performance or phonogram in digital form or other electronic medium.

**2. Fair Dealing Rights.** Nothing in this License is intended to reduce, limit, or restrict any uses free from copyright or rights arising from limitations or exceptions that are provided for in connection with the copyright protection under copyright law or other applicable laws.

**3. License Grant.** Subject to the terms and conditions of this License, Licensor hereby grants You a worldwide, royalty-free, non-exclusive, perpetual (for the duration of the applicable copyright) license to exercise the rights in the Work as stated below:
- a) to Reproduce the Work, to incorporate the Work into one or more Collections, and to Reproduce the Work as incorporated in the Collections;
- b) to create and Reproduce Adaptations provided that any such Adaptation, including any translation in any medium, takes reasonable steps to clearly label, demarcate or otherwise identify that changes were made to the original Work. For example, a translation could be marked "The original work was translated from English to Spanish," or a modification could indicate "The original work has been modified.";
- c) to Distribute and Publicly Perform the Work including as incorporated in Collections; and,
- d) to Distribute and Publicly Perform Adaptations.

The above rights may be exercised in all media and formats whether now known or hereafter devised. The above rights include the right to make such modifications as are technically necessary to exercise the rights in other media and formats. Subject to Section 8(f), all rights not expressly granted by Licensor are hereby reserved, including but not limited to the rights set forth in Section 4(d).

**4. Restrictions.** The license granted in Section 3 above is expressly made subject to and limited by the following restrictions:
- a) You may Distribute or Publicly Perform the Work only under the terms of this License. You must include a copy of, or the Uniform Resource Identifier (URI) for, this License with every copy of the Work You Distribute or Publicly Perform. You may not offer or impose any terms on the Work that restrict the terms of this License or the ability of the recipient of the Work to exercise the rights granted to that recipient under the terms of the License. You may not sublicense the Work. You must keep intact all notices that refer to this License and to the disclaimer of warranties with every copy of the Work You Distribute or Publicly Perform. When You Distribute or Publicly Perform the Work, You may not impose any effective technological measures on the Work that restrict the ability of a recipient of the Work from You to exercise the rights granted to that recipient under the terms of the License. This Section 4(a) applies to the Work as incorporated in a Collection, but this does not require the Collection apart from the Work itself to be made subject to the terms of this License. If You create a Collection, upon notice from any Licensor You must, to the extent practicable, remove from the Collection any credit as required by Section

      granted under this License if Your exercise of such rights is for a purpose or use which is otherwise than noncommercial as permitted under Section 4(b) and otherwise waives the right to collect royalties through any statutory or compulsory licensing scheme; and,

   iii. f.Voluntary License Schemes. The Licensor reserves the right to collect royalties, whether individually or, in the event that the Licensor is a member of a collecting society that administers voluntary licensing schemes, via that society, from any exercise by You of the rights granted under this License that is for a purpose or use which is otherwise than noncommercial as permitted under Section 4(c).

e) Except as otherwise agreed in writing by the Licensor or as may be otherwise permitted by applicable law, if You Reproduce, Distribute or Publicly Perform the Work either by itself or as part of any Adaptations or Collections, You must not distort, mutilate, modify or take other derogatory action in relation to the Work which would be prejudicial to the Original Author's honor or reputation. Licensor agrees that in those jurisdictions (e.g. Japan), in which any exercise of the right granted in Section 3(b) of this License (the right to make Adaptations) would be deemed to be a distortion, mutilation, modification or other derogatory action prejudicial to the Original Author's honor and reputation, the Licensor will waive or not assert, as appropriate, this Section, to the fullest extent permitted by the applicable national law, to enable You to reasonably exercise Your right under Section 3(b) of this License (right to make Adaptations) but not otherwise.

## 5. Representations, Warranties and Disclaimer

UNLESS OTHERWISE MUTUALLY AGREED TO BY THE PARTIES IN WRITING, LICENSOR OFFERS THE WORK AS-IS AND MAKES NO REPRESENTATIONS OR WARRANTIES OF ANY KIND CONCERNING THE WORK, EXPRESS, IMPLIED, STATUTORY OR OTHERWISE, INCLUDING, WITHOUT LIMITATION, WARRANTIES OF TITLE, MERCHANTIBILITY, FITNESS FOR A PARTICULAR PURPOSE, NONINFRINGEMENT, OR THE ABSENCE OF LATENT OR OTHER DEFECTS, ACCURACY, OR THE PRESENCE OF ABSENCE OF ERRORS, WHETHER OR NOT DISCOVERABLE. SOME JURISDICTIONS DO NOT ALLOW THE EXCLUSION OF IMPLIED WARRANTIES, SO SUCH EXCLUSION MAY NOT APPLY TO YOU.

**6. Limitation on Liability.** EXCEPT TO THE EXTENT REQUIRED BY APPLICABLE LAW, IN NO EVENT WILL LICENSOR BE LIABLE TO YOU ON ANY LEGAL THEORY FOR ANY SPECIAL, INCIDENTAL, CONSEQUENTIAL, PUNITIVE OR EXEMPLARY DAMAGES ARISING OUT OF THIS LICENSE OR THE USE OF THE WORK, EVEN IF LICENSOR HAS BEEN ADVISED OF THE POSSIBILITY OF SUCH DAMAGES.

## 7. Termination

a) This License and the rights granted hereunder will terminate automatically upon any breach by You of the terms of this License. Individuals or entities

    who have received Adaptations or Collections from You under this License, however, will not have their licenses terminated provided such individuals or entities remain in full compliance with those licenses. Sections 1, 2, 5, 6, 7, and 8 will survive any termination of this License.

b) Subject to the above terms and conditions, the license granted here is perpetual (for the duration of the applicable copyright in the Work). Notwithstanding the above, Licensor reserves the right to release the Work under different license terms or to stop distributing the Work at any time; provided, however that any such election will not serve to withdraw this License (or any other license that has been, or is required to be, granted under the terms of this License), and this License will continue in full force and effect unless terminated as stated above.

## 8. Miscellaneous

a) Each time You Distribute or Publicly Perform the Work or a Collection, the Licensor offers to the recipient a license to the Work on the same terms and conditions as the license granted to You under this License.

b) Each time You Distribute or Publicly Perform an Adaptation, Licensor offers to the recipient a license to the original Work on the same terms and conditions as the license granted to You under this License.

c) If any provision of this License is invalid or unenforceable under applicable law, it shall not affect the validity or enforceability of the remainder of the terms of this License, and without further action by the parties to this agreement, such provision shall be reformed to the minimum extent necessary to make such provision valid and enforceable.

d) No term or provision of this License shall be deemed waived and no breach consented to unless such waiver or consent shall be in writing and signed by the party to be charged with such waiver or consent.

e) This License constitutes the entire agreement between the parties with respect to the Work licensed here. There are no understandings, agreements or representations with respect to the Work not specified here. Licensor shall not be bound by any additional provisions that may appear in any communication from You. This License may not be modified without the mutual written agreement of the Licensor and You.

f) The rights granted under, and the subject matter referenced, in this License were drafted utilizing the terminology of the Berne Convention for the Protection of Literary and Artistic Works (as amended on September 28, 1979), the Rome Convention of 1961, the WIPO Copyright Treaty of 1996, the WIPO Performances and Phonograms Treaty of 1996 and the Universal Copyright Convention (as revised on July 24, 1971). These rights and subject matter take effect in the relevant jurisdiction in which the License terms are sought to be enforced according to the corresponding provisions of the implementation of those treaty provisions in the applicable national law. If the standard suite of rights granted under applicable copyright law includes additional rights not granted under this License, such additional rights are deemed to be included in the License; this License is not intended to restrict the license of any rights under applicable law.

**Creative Commons Notice**

Creative Commons is not a party to this License, and makes no warranty whatsoever in connection with the Work. Creative Commons will not be liable to You or any party on any legal theory for any damages whatsoever, including without limitation any general, special, incidental or consequential damages arising in connection to this license. Notwithstanding the foregoing two (2) sentences, if Creative Commons has expressly identified itself as the Licensor hereunder, it shall have all rights and obligations of Licensor.

Except for the limited purpose of indicating to the public that the Work is licensed under the CCPL, Creative Commons does not authorize the use by either party of the trademark "Creative Commons" or any related trademark or logo of Creative Commons without the prior written consent of Creative Commons. Any permitted use will be in compliance with Creative Commons' then-current trademark usage guidelines, as may be published on its website or otherwise made available upon request from time to time. For the avoidance of doubt, this trademark restriction does not form part of the License.

Creative Commons may be contacted at http://creativecommons.org/.

**Publishing studies** series

Community Informatics (CI) is the application of information and communications technologies (ICTs) to enable community processes and the achievement of community objectives. CI goes beyond the "Digital Divide" to making ICT access usable and useful to excluded populations and communities for local economic development, social justice, and political empowerment. CI approaches ICTs from a "community" perspective and develops strategies and techniques for managing their use by communities both virtual and physical including the variety of Community Networking applications. CI assumes that both communities have characteristics, requirements, and opportunities that require different strategies for ICT intervention and development from individual access and use. Also, CI addresses ICT use in Developing Countries as well as among the poor, the marginalized, the elderly, or those living in remote locations in Developed Countries. CI is of interest both to ICT practitioners and academic researchers and addresses the connections between the policy and pragmatic issues arising from the tens of thousands of Community Networks, Community Technology Centres, Telecentres, Community Communications Centres, and Telecottages globally along with the rapidly emerging field of electronically based virtual "communities".

Michael Gurstein, Ph.D. is Executive Director of the Centre for Community Informatics Research, Development and Training (Vancouver BC), a Director of The Information Society Institute, Cape Peninsula University of Technology, Cape Town South Africa; and Research Professor in the School of Computer and Information Systems at the New Jersey Institute of Technology, Newark. His seminal book "Community Informatics: Enabling Communities with Information and Communications Technologies" (Idea Group, 2000) spurred the emergence of Community Informatics as the discipline underpinning the social appropriation of ICT. Dr. Gurstein has served on the Board of the Vancouver Community Network, the British Columbia Community Networking Association, Telecommunities Canada, and was a Charter Member of the Steering Committee of the Global Community Networking Partnership. He is the Editor in Chief of the Journal of Community Informatics and was Foundation Chair of the Community Informatics Research Network.

ISSN 1973-6061

ISBN 978-88-7699-097-7

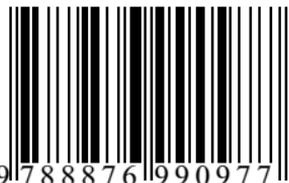